\documentclass[12pt]{article}
\pdfoutput=1
\usepackage[utf8x]{inputenc}
\usepackage{amssymb,amsmath,mathrsfs,enumerate}
\usepackage{rotate,multicol}
\usepackage{graphicx,wrapfig}
\usepackage{tikz,graphics,color,fullpage,float,epsf,caption,subcaption}
\usepackage{soul}
\usepackage{float}
\usepackage{cite}
\usepackage[colorlinks=true,
  linkcolor=red,
  urlcolor=blue,
  citecolor=blue]{hyperref}
\usepackage{multirow}
\usepackage{color}
\usepackage{cite}
\usepackage{comment}
\usepackage{cancel}
\usepackage{eurosym}
\usepackage{hyperref} 
   
\long\def\rpl#1!!#2!!{\textcolor{red}{#1} \textcolor{blue}{#2}}

\def \order(#1){{\cal O} \left(#1 \right)}

\newmuskip\pFqskip
\mathchardef\pFcomma=\mathcode`, 

\evensidemargin=\oddsidemargin
\allowdisplaybreaks

\begin{document}

\begin{flushright}
 BONN-TH-2023-05
\end{flushright}


\begin{center}
  {\Large \bf  Light long-lived particles at the FCC-hh with the proposal for a dedicated forward detector FOREHUNT and a transverse detector DELIGHT } \\\vspace*{1cm}  
\renewcommand{\thefootnote}{\fnsymbol{footnote}}  
{{\sf Biplob Bhattacherjee$^1$\footnote{biplob@iisc.ac.in}},  
{\sf  ~Herbi K. Dreiner$^2$\footnote{dreiner@uni-bonn.de}},
{\sf ~Nivedita Ghosh$^1$\footnote{ niveditag@iisc.ac.in (Corresponding Author)}}, 
{\sf ~Shigeki Matsumoto$^3$\footnote{shigeki.matsumoto@ipmu.jp}},\\
{\sf Rhitaja Sengupta$^2$\footnote{rsengupt@uni-bonn.de}}, 
{\sf Prabhat Solanki$^1$\footnote{prabhats@iisc.ac.in }}
}\\
\vspace{10pt}{\small} {$^1$Centre for High Energy Physics, Indian Institute of Science, Bengaluru 560012, India \\
$^2$ Bethe Center for Theoretical Physics and Physikalisches Institut der Universit\"at Bonn, \\
Nußallee 12, 53115 Bonn, Germany \\
$^3$ Kavli IPMU (WPI), UTIAS, University of Tokyo, Kashiwa, Chiba, 277-8583, Japan}
 
\end{center}

\renewcommand*{\thefootnote}{\arabic{footnote}}
\setcounter{footnote}{0} 

\bigskip
\begin{abstract}
In this paper, we propose a dedicated forward detector, FOREHUNT (FORward~Experiment~for~HUNdred~TeV), for 100\,TeV FCC-hh for the detection of light long-lived particles (LLP) coming from $B$-meson decay. We calculate the signal acceptance as a function of mass and proper decay length of the LLP for 100\,TeV and interpret our result in terms of model parameters for models of dark Higgs scalar and heavy neutral leptons.
We also compare the sensitivity with proposed transverse detectors like MATHUSLA, CODEX-b for HL-LHC, and DELIGHT (Detector for long-lived particles at high energy of 100\,TeV) for FCC-hh. Our analysis reveals that if the LLP is light ($\lesssim 4.4$\,GeV) and has a low proper decay length ($<10$~m), a forward detector like FOREHUNT is the best option to look for the decaying LLP, while DELIGHT is preferable for higher proper decay lengths.

\end{abstract}

\section{Introduction}
\label{s:intro}

Our searches for physics beyond the Standard Model (BSM) have yet to reveal any conclusive hint of its existence. Under these circumstances, we are compelled to revisit the assumptions behind these searches and ensure that we give equal attention to all possible forms of new physics. In this regard, a recently growing area of interest is the study of signatures of long-lived BSM particles (LLPs), which have macroscopic proper decay lengths ($c\tau\gtrsim\mathcal{O}({\rm mm})$) in the collider detectors. 
Several BSM physics scenarios, like SUSY\,\cite{Dimopoulos:1996vz,Farrar:1996rg,Giudice:1998bp,Baer:1998pg,Choudhury:1999tn,Mafi:1999dg,Kraan:2004tz, Mackeprang:2006gx, Arkani-Hamed:2004ymt,Giudice:2004tc,Hewett:2004nw,Arvanitaki:2005nq,Dreiner:2009ic,Meade:2010ji, Fan:2012jf, Bhattacharyya:2012ct, Bhattacherjee:2012ed, Arvanitaki:2012ps, Banerjee:2016uyt, Nagata:2017gci, Dercks:2018wum, Banerjee:2018uut, Ito:2018asa, Dreiner:2020qbi}, hidden valley models~\cite{Strassler:2006im, Strassler:2006ri}, and models involving axion-like particles (ALPs)~\cite{PhysRevLett.38.1440, PhysRevD.16.1791, Jaeckel:2010ni, Bauer:2017ris} or dark photons ~\cite{HOLDOM1986196,Bauer:2018onh,Fabbrichesi:2020wbt,Caputo:2021eaa}, predict LLPs.
The signatures of LLPs depend on where these new particles decay, in addition to their decay modes. The decay length of a particle in the collider detector ($d$) is a product of its boost factor, $\beta\gamma$, and its proper decay length, $c\tau$, where the former depends on the mass of the particle and its production mode. 
A multitude of searches have been designed and performed for various kinds of LLPs in experiments at the Large Hadron Collider (LHC)\,\cite{deVries:2015mfw,ATLAS:2015xit,ATLAS:2018rjc,ATLAS:2018niw,ATLAS:2018tup,ATLAS:2019fwx,ATLAS:2019tkk,ATLAS:2019jcm,ATL-PHYS-PUB-2019-002,ATLAS:2020xyo,ATLAS-CONF-2021-032,ATLAS:2021jig,CMS:2014hka,CMS:2017kku,CMS:2018bvr,CMS-PAS-FTR-18-002,CMS:2019zxa,CMS:2020atg,CMS-PAS-EXO-19-021,CMS:2021juv,CMS:2021kdm,CMS:2021yhb,LHCb:2016buh,LHCb:2016inz,LHCb:2017xxn,LHCb:2019vmc,LHCb:2020akw} for different signatures, like displaced vertices, non-pointing photons, delayed objects, or clusters of hits in the muon spectrometer of the detector, without any activity in the inner detectors. The phenomenology of LLPs has also been studied in low energy $e^+$-$e^-$ colliders, like Belle-II\,\cite{Batell:2009yf,Belle-II:2010dht,Belle-II:2018jsg,Duerr:2019dmv,Filimonova:2019tuy,Zhou:2021ylt,Dreyer:2021aqd,Bertholet:2021hjl,Ferber:2022rsf,Bandyopadhyay:2022klg} and neutrino facilities\,\cite{Batell:2009di,Dev:2021qjj,Batell:2022xau,Coloma:2023adi,Batell:2023mdn}. 

Searches for heavier LLPs ($\gtrsim\mathcal{O}(100{\rm~GeV})$) are more straightforward than for lighter ones ($\lesssim\mathcal{O}(10{\rm~GeV})$). In spite of the smaller cross-sections, heavy BSM particles' decay products are energetic, making it easier to trigger them. However, for heavy LLPs, the decay products can be significantly delayed if the decay is highly displaced. In such situations, although standard triggers might not be efficient, dedicated triggers for delayed jets can prove effective, as shown in Refs.\,\cite{Bhattacherjee:2020nno,CERN-LHCC-2020-004,Bhattacherjee:2021qaa}.
Heavy LLPs usually have less boost, leading to smaller decay lengths in the detector; hence, better reconstruction of the displaced vertex is possible.
Less boost also implies that these massive LLPs travel slower and cause a high time delay of their decay products as compared to prompt relativistic particles.
The production of heavy particles is also accompanied by more radiation which aids in reconstructing the correct primary vertex (PV) in the collision from where the heavy particle has been produced. This becomes important in pile-up (PU) mitigation as well as in estimating the correct time delay associated with the decay of the LLP.
Collider detectors, like calorimeters, are segmented in the $\eta$ and $\phi$ directions with respect to the primary vertex. Particles originating near the primary vertex with $\eta_p$,$\phi_p$ are detected in the detector segment corresponding to $\eta_p$,$\phi_p$, since their direction is aligned to these segments.
However, the displaced decay products originate from the secondary vertex. If they are produced in the $\eta_d$,$\phi_d$ direction, they are not necessarily detected in the detector segment corresponding to $\eta_d$,$\phi_d$, since their direction is not aligned with the direction of the detector segments. This creates a mismatch between the displaced particle's actual direction and the segment where it gets detected.
Due to the layered structure and projective geometry of the calorimeter, this mismatch results in more spread-out energy deposits of displaced particles than that of prompt particles.
For lighter LLPs, the decay products are boosted in the direction of the mother particle, and thus, this mismatch is smaller than for heavier LLPs. 
Therefore, the decay products of heavier LLPs have more
mismatch with the detector’s $\eta$-$\phi$ directions $-$ 
leading to calorimeter deposits that appear elongated in the $\eta$-$\phi$ plane\,\cite{CMS:2012bbi,Bhattacherjee:2019fpt}.
These make the signatures of LLPs with large masses substantially different from the SM processes.
For light LLPs, the search becomes more challenging.
The first difficulty is disentangling the backgrounds from the known long-lived light particles of the SM, like $K_S^0$ or $\Lambda$. 
Moreover, events with light LLPs are difficult to trigger in the absence of any associated prompt hard objects, and the chances of incorrectly assigning the PV also increase.
Owing to high boosts, lighter LLPs have larger decay lengths in the detector, making the reconstruction of the secondary vertex (SV) difficult.
Being mostly relativistic, their decay products also do not have significant time delay as compared to the SM processes. 
In the present work, we, therefore, take a closer look at light LLPs.

The main detectors of the LHC experiment, CMS, ATLAS, and LHCb, have performed some dedicated searches for such light LLPs\,\cite{CMS:2021sch,ATLAS:2022izj,LHCb:2016awg,LHCb:2017trq}.
The CMS and ATLAS detectors extend up to $\mathcal{O}$(10\,m), and lose sensitivity to BSM particles which have higher proper decay lengths. 
Light LLPs with large proper decay lengths are more likely to decay outside the LHC main detector volumes due to the high boost values.
Dedicated LLP detectors can play a crucial role in probing such highly displaced scenarios.
Depending on the production mode, the direction of the LLP boost can be in the transverse or forward direction. For example, if the LLP is produced from the decay of a Higgs boson, light LLPs are more likely to be produced in the transverse direction. In contrast, if they are produced from meson decays, they are boosted in the forward direction. We focus on the latter possibility in the present work, mainly considering LLPs from the $B$ meson decay due to the larger available phase space.
At LHC, FASER (ForwArd Search ExpeRiment)~\cite{Feng:2017uoz, Boiarska:2019vid} is one such detector specifically designed to look for light LLPs in the far forward region. FASER is a cylindrical detector placed 480\,m away from the ATLAS interaction point (IP) and is expected to collect 150\,$\mathrm{fb^{-1}}$ of data in Run-3 of LHC. At the high luminosity LHC (HL-LHC), FASER will be upgraded to FASER 2 with a much bigger detector volume and is designed to collect 3\,$\mathrm{ab^{-1}}$ of data. 
Other proposals of forward LLP detectors include the MAPP-MoEDAL\,\cite{Pinfold:2017dot,MoEDAL-MAPP:2022kyr} and the Forward Multi-particle Spectrometer~\cite{Albrow:2020duu}.
There are several other proposed experiments in the transverse direction, like MATHUSLA~\cite{Chou:2016lxi,Curtin:2018ees,Curtin:2018mvb,Alpigiani:2020iam},
CODEX-b~\cite{Gligorov:2017nwh,Aielli:2019ivi}, and ANUBIS~\cite{Bauer:2019vqk}. These transverse detectors also have some sensitivity to probe light LLPs from meson decays. 
Previous beam dump experiments, like the NuTeV and BEBC, also provide bounds on such light particles\,\cite{Dedes:2001zia,Marocco:2020dqu,Barouki:2022bkt}, while the
proposed and ongoing beam dump experiments, like SHiP~\cite{Bonivento:2013jag,Alekhin:2015byh,SHiP:2015vad}, 
LongQuest~\cite{Tsai:2019buq}, 
NA62\,\cite{Lanfranchi:2671026}, 
NA64~\cite{NA64:2016oww,Gninenko:2019qiv,Banerjee:2019pds,NA64:2020qwq}, and DarkQuest~\cite{Berlin:2018pwi,Batell:2020vqn,Blinov:2021say,Apyan:2022tsd,Tsai:2019buq} are also sensitive to light LLPs.

In this paper, we study the enhancement in the sensitivity of light LLPs from meson decays for the 100\,TeV future circular collider, FCC-hh. A number of dedicated LLP detectors have been proposed for the future lepton collider, FCC-ee\,\cite{Chrzaszcz:2020emg,Schafer:2022shi}, and a transverse LLP detector, DELIGHT, has been proposed for the FCC-hh\,\cite{Bhattacherjee:2021rml}\,\footnote{Another proposal for a dedicated LLP detector at the FCC-hh optimized for heavy neutral leptons can be found in Ref.\,\cite{Boyarsky:2022epg}.}. 
Since the proposed 100\,TeV collider is still in its early stages of design and construction, there is an opportunity to design and propose new forward detectors that can efficiently probe such light particles with displaced decays. 
Unlike limited by the available spaces around the HL-LHC facility, the future colliders, therefore, provide more room for selecting optimal designs and locations for the LLP detectors, which can improve our sensitivity to various new physics models.
For the present work, we study two BSM scenarios that give rise to mesons decaying into light LLPs. 
First, we study a simple extension of the SM by adding a long-lived scalar, $\phi$, that couples to the discovered Higgs boson with a mixing angle, $\theta$. This kind of scenario is well-motivated from various BSM perspectives, like producing the correct relic density of dark matter (DM), solving the small-scale problems in structure formation, or in theories of inflation\,\cite{Patt:2006fw,Feng:2008ya,Tulin:2017ara,Shaposhnikov:2006xi,Bezrukov:2009yw,Bezrukov:2013fca,Bramante:2016yju}. Such long-lived scalars can be produced at colliders via the decay of $B$-mesons in association with SM particles.
Secondly, we consider heavy neutral leptons (HNLs)\,\cite{Keung:1983uu,Ferrari:2000sp,Helo:2013esa,Maiezza:2015lza,Izaguirre:2015pga,Batell:2016zod,Nemevsek:2016enw,Accomando:2017qcs,Helo:2018qej,Chakraborty:2018khw,DeVries:2020jbs,Abdullahi:2022jlv,Batell:2022ogj}, which can be long-lived in various regions of the parameter space, and if they are light enough, they can also be produced from meson decays.
We propose various viable detector configurations in the forward direction for different LLP benchmarks from these models after optimizing their positions and dimensions.
We also compare the results with other potential transverse LLP detectors, such as CODEX-b and MATHUSLA for HL-LHC, and DELIGHT for FCC-hh, to identify regions where these have complementary or overlapping sensitivities.  
Furthermore, we examine the possibility of placing multiple detectors in the forward and slightly off-axis direction to evaluate their impact on signal efficiency for our benchmark points.

The rest of the paper is organized as follows: In Section~\ref{s:model}, we describe the signal models under investigation, including the production modes and decay modes of the LLPs. We discuss our analysis setup and its validation against existing results in Section~\ref{s:analysis}. 
In Section~\ref{s:result}, we propose and study the sensitivity of forward dedicated LLP detectors at the 100\,TeV collider, and optimize the positions, dimensions, and number of detectors to enhance the performance for our LLP benchmarks, keeping the design feasible.
We also compare the sensitivity of our proposed forward detectors with the existing proposal of DELIGHT, a transverse detector for the FCC-hh.
Finally, in Section~\ref{s:summary}, we present a discussion of our findings and conclude with the summary.


\section{Model}
\label{s:model}

In this work, we consider two well-established models for LLPs, namely the dark Higgs boson and heavy neutral leptons.
In the first scenario, we extend the SM minimally with a new singlet scalar particle $\Phi$ with a quartic coupling to the SM Higgs doublet ($H$).
The corresponding Lagrangian can be written as~\cite{Krnjaic:2015mbs,Li:2022zgr}:
\begin{eqnarray}
{\cal L} = {\cal L}_{\rm SM} + {\mu^2_{\Phi}} \Phi^2 - \frac{1}{4}\lambda_{\Phi} \Phi^4 - \epsilon \Phi^2|H|^2 \,.
\label{eq:lag}
\end{eqnarray} 

where $ {\cal L}_{\rm SM}$ is the Lagrangian for the SM particles.
The last term in Eq.~(\ref{eq:lag}) introduces the mixing between the SM Higgs and the new scalar $\Phi$. After electroweak symmetry breaking,
the scalar $\Phi$ and the real neutral component of $H$, namely $h$, get $\it{vaccum~ expectation~ values}$ (vevs). Rotating them from gauge basis to mass basis, one gets the physical states as follows:
\begin{eqnarray}
 h_{125}= \Phi \sin \theta + h \cos \theta\,, \\ \nonumber
 \phi = \Phi \cos \theta - h \sin \theta ,
\end{eqnarray}
where $h_{125}$ resembles the 
discovered Higgs boson and $\phi$ is called the dark Higgs scalar. The new scalar, $\phi$, 
couples with the fermions via a mixing angle, $\theta$, with the Higgs boson~\cite{Krnjaic:2015mbs} as follows:
\begin{eqnarray}
 {\cal L}_{\rm int} =   \phi  \sin\theta  \sum_f   \frac{m_f}{v}   \bar f f \,,~~
 \label{eq:lagphi}
\end{eqnarray} 
where $f$ denotes the SM fermions and $v\simeq 246$\,GeV.

The decay of the scalar $\phi$ to fermions is given at tree-level by~\cite{Krnjaic:2015mbs,Winkler:2018qyg,FASER:2018eoc}
\begin{eqnarray}
\Gamma_{\! \phi \to f \bar{f}} = \frac{N_c G_F m_{\phi} m_f^2 \sin^2\theta}{4\sqrt{2}\pi} \left(1-\frac{4 m_f^2} {m_{\phi}^2}\right)^{3/2}
\end{eqnarray}
where $G_F$ is the Fermi constant and $N_c$ is the color quantum number. Depending on its mass, the light
scalar can also decay 
into a pair of gluons, pions, kaons, and photons. If the mixing angle, $\sin\theta$, is very small, the proper decay length of the dark Higgs 
scalar is $c\tau \sim $ few mm, making the particle long-lived. For example, for a mass $m_{\phi} = 1.0$\,GeV, one requires $\sin\theta < 
{\cal{O}}(10^{-4})$, to make $\phi$ long-lived.

In this paper, we assume $\phi$ to be very light, with its mass ranging from $\sim$ 100\,MeV to few GeV. Such light scalars can be produced from various meson decays. We mainly focus on the production of $\phi$ in the $B^{\pm} \to K^{\pm} \phi$ mode~\cite{Araki:2020wkq,Li:2022zgr}. 
We have chosen this particular decay mode due to its accessibility to probe a larger range of dark scalar masses.
The partial width of $B^\pm$ in this mode is given by\,\cite{Dolan:2014ska}
\begin{eqnarray} \label{eq:fullbwidth} \hspace{-0.1cm} &&
\Gamma_{\! B^{\pm} \to K^{\pm} \phi} = \frac{|C_{sb}|^2  f_0(m_{\phi}^2)^2 \!}{16 \pi m_{B^+}^3}\! \left( \! \frac{m_{B^+}^2 - m_{K^+}^2}{m_b - m_s} \!\! \right)^2  \hspace{-0.2cm} \xi(m_{B^+}, m_{K^+}, m_{\phi})  ,~~~~~~~
\\ \nonumber \\
&&   ~~~~~~~~~~~~~~~\xi(a,b,c) \,\equiv\, \sqrt{   (a^2 - b^2 -c^2)^2 - 4 b^2 c^2 }~,
\end{eqnarray}
where the scalar form factor, $f_0$, can be parametrized as $f_{0}(q^2) = \frac{0.33}{1 - (\frac{q^{2}}{38 \,{\rm GeV}^2})}$\,\cite{Ball:2004ye}, and,
\begin{equation}
C_{sb} \,\equiv\,  \frac{3 g_W^2 m_{b} m_t^2 V_{ts}^* V_{tb}   \sin\theta  }{ 64 \pi^2 m_W^2 v }  = 6.4 \times 10^{-6} \sin\theta ~.~~~~
\end{equation}

The total $B$-meson width is, $\Gamma_{B^\pm} = 4.1 \times 10^{-13}$\,GeV\,\cite{Agashe:2014kda}, so the branching ratio has the approximate scaling
\begin{eqnarray}
{\cal B}r(B^\pm \to \! K^\pm \phi) = \frac{\Gamma_{\! B^{\pm} \to K^{\pm} \Phi}}{\Gamma_{B^\pm}} 
\end{eqnarray}
In order to reduce the theoretical uncertainties arising from the form factors, we consider the inclusive branching fraction of $B$ mesons to any strange meson, $X_s$, and the dark Higgs boson\,\cite{Chivukula:1988gp,Dolan:2014ska,Feng:2017vli,FASER:2018eoc}.
The inclusive branching fraction of the $B$ mesons to the scalar LLP is then given by\,\cite{Chivukula:1988gp,Feng:2017vli}:
\begin{eqnarray}
 \text{Br}(B \to X_S \phi) = 5.7~ \text{sin}^2\,\theta\,\left(1-\frac{m_{\phi}^2}{m_b^2}\right)^2,
 \label{eq:br_BtoKphi}
\end{eqnarray}
where $m_b=4.75$\,GeV. 

Another possible production mode of $\phi$ can be from the decay of kaons~\cite{Araki:2020wkq}, however, the ${\cal B}r(K^\pm \to \pi^\pm \phi)  \sim  
2.77\times10^{-2}  \xi( m_{K}, m_{\pi}, m_\phi)  \sin^2\!\theta $ is much smaller than ${\cal B}r(B^\pm \to \! K^\pm \phi)$ and is only allowed in a narrow 
mass range when $m_{\phi} < 0.3$ GeV. It is also severely constrained by other current experiments like E949~\cite{BNL-E949:2009dza}. 
Additionally, since the kaons are also highly displaced ($c\tau=3.714$\,m), it reduces the probability of containing the LLP decay within smaller regions (see Appendix\,\ref{app:kaon}). As a result, although the overall kaon production is much higher than the $B$-meson production, it would not have a significant effect on the sensitivity.
The branching of the $D$-meson decaying to scalars is much more suppressed due to the absence of a top quark in the loop~\cite{Li:2022zgr}. 
Therefore, in the present work, we have focused on the production from $B$-meson decays. Contributions from the other mesons will contribute to the 
signal constructively, hence making our analysis conservative. 
 
In the second scenario, we have considered an extension of the SM with HNLs~\cite{PhysRevLett.40.1688,Kling:2018wct,FASER:2018eoc,Abdullahi:2022jlv}. 
In this model, we extend the SM Lagrangian by adding the following term:
\begin{equation}
{\cal L}_{\rm int} = -\sum_I y_{\alpha I}(\bar{L}_{\alpha} H) N_I\,,
\label{eq:lagHNL}
\end{equation}

where the Yukawa couplings are given by $y_{\alpha I}$, ${L}_{\alpha}$ are the SM lepton doublets, $H$ is the Higgs doublet, and the HNL fields are 
denoted by $N_I$. The index $\alpha$ runs over the three generations of SM leptons, while the index $I$ runs over the different HNL fields. The HNLs can have both Majorana and Dirac mass terms.
After EWSB, there will be a mixing between the SM leptons and the HNLs, and we denote the mixing matrix element by $|U_{N{\alpha}}|$. 
In principle, there will be a 
mixing of the HNLs with all three generations of SM leptons. However, for electrons and muons, the mixing $|U_{N{\alpha}}|$ is highly constrained 
by current bounds~\cite{Drewes:2018irr}, unlike the case for $\tau$ leptons. We, therefore, consider only mixing of the HNL with the $\tau$ leptons in 
the present work, and we denote the HNL mass as $m_N$ hereafter. 

We assume the following production modes of the HNL coming from the $B$-meson decay~\cite{Kling:2018wct}\,\footnote{The HNLs can also come from the prompt $W$ decay. However, the $W^\pm$ production cross-section is around $1.3\times10^9$\,fb\,\cite{Mangano:2016jyj} at 100\,TeV, which is at least three orders of magnitude less than the $B$-meson production rate at FCC-hh. Therefore, this will lead to a subdominant contribution to the sensitivity and we do not consider it in the present study.}:
\begin{eqnarray}
B^0 &\rightarrow& D^{\pm} \tau^{\mp} N_{\tau} \,, \label{eq:hnl_a} \\
B^{\pm}&\rightarrow& D^0 \tau^{\pm} N_{\tau}  \,, \label{eq:hnl_b} \\
B^{\pm}&\rightarrow& \tau^{\pm} N_{\tau}  \,.\label{eq:hnl_c}
\label{eq:hnlproduce}
\end{eqnarray}

The production of HNLs is phenomenologically different than the dark Higgs scalars, as the former can be produced via both the two-body 
and three-body decays of $B$-mesons. The branching ratios of the processes described in Eqs.~(\ref{eq:hnl_a})-(\ref{eq:hnl_c}) can be obtained from 
Ref.~\cite{Gorbunov:2007ak,Graverini:2133817}. 
In the following section, we describe our analysis setup and present its validation.

\section{Analysis Setup and Validation}
\label{s:analysis}

In order to detect light and weakly interacting particles, a dedicated forward detector, called FASER has been introduced to operate at 
the LHC while the collision energy is $\sqrt{s}=13.6$ TeV. Additionally, an upgraded version of the 
detector, FASER2 \cite{FASER:2018eoc}, has been proposed specifically for the 
HL-LHC, which is expected to run at a center-of-mass energy of 14\,TeV. FASER is designed with a radius\,($R$) of 10\,cm and length\,($L_d$) of 1.5\,m. It is placed at 480\,m in the $z$-axis from the 
IP, aiming to collect data during LHC run3 2021-2023~\cite{FASER:2018eoc} for integrated luminosity $150 ~\rm{fb^{-1}}$. It is envisaged 
that FASER2 will possess a radius of 1\,m and a length of 5\,m,
and it is expected to collect data for HL-LHC in the era 2026-2035 with an integrated luminosity of ${\cal L} =3000 ~\rm {fb^{-1}}$.  

Let us now discuss the setup used for our analysis. To ensure a consistent simulation setup, we reproduce the proposed sensitivity of 
FASER2, focusing on the dark Higgs boson model discussed in Section~\ref{s:model}, which matches with benchmark scenario S1 in 
Ref.~\cite{Li:2022zgr}. For our simulation, we generate pairs of bottom quarks in $pp$ collisions at a center-of-mass energy of 
$\sqrt{s}=$ 14\,TeV using \texttt{MadGraph5-aMC@NLO}~\cite{Alwall:2014hca,Frederix:2018nkq}, 
with the \texttt{nn23nlo} parton distribution function (PDF) at next-to-leading order (NLO). The resulting output from \texttt{MadGraph} 
is then processed through \texttt{pythia8}~\cite{Bierlich:2022pfr}, which handles the hadronization of the $b$-quarks into $B$-mesons 
and their subsequent decay into a kaon and a dark Higgs boson. To incorporate the new decay modes of the $B$-mesons to LLPs, the 
corresponding decay table in \texttt{pythia8} has been modified accordingly, and the lifetime of the long-lived particle has been explicitly 
adjusted within the simulation framework.
 
\begin{figure}[htb!]
  \centering  
   \includegraphics[width=0.75\textwidth]{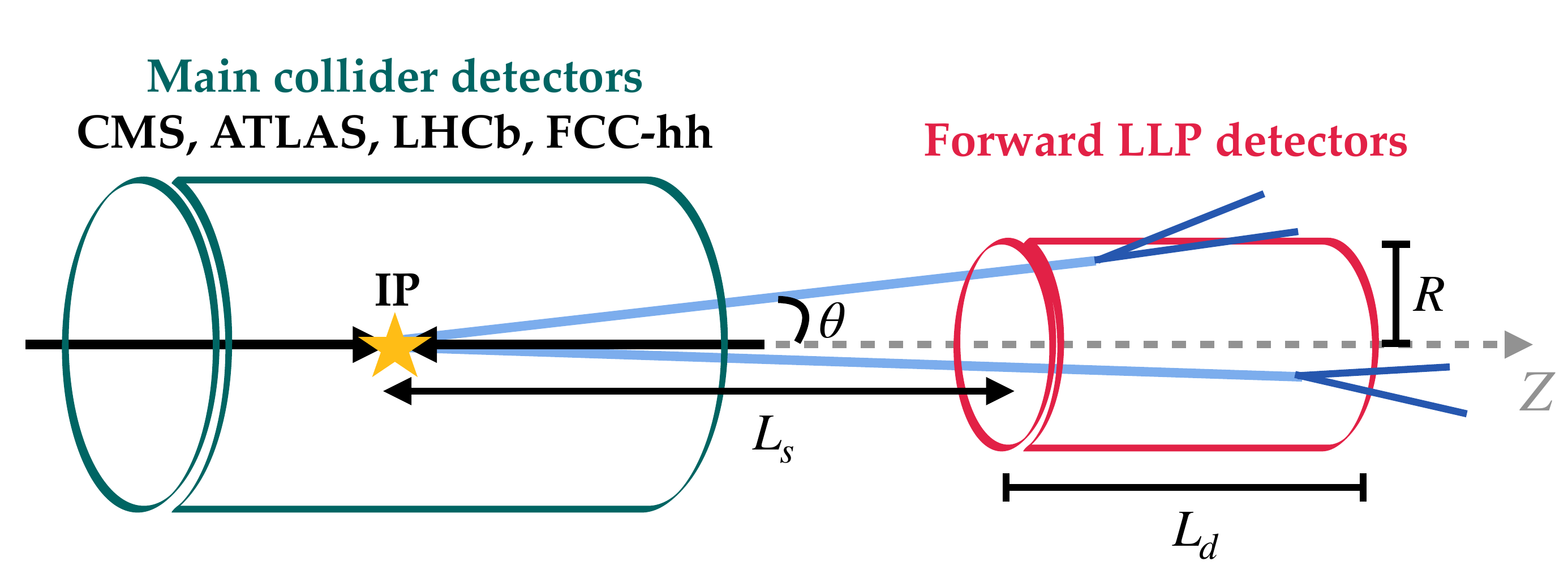}
   \caption{A schematic diagram of forward dedicated LLP detectors.}
   \label{fig:forward_detector}
\end{figure}

The probability of a long-lived particle with mean proper decay length $c\tau$ to decay inside a forward detector can be calculated using~\cite{Dercks:2018eua}
\begin{equation}
P_{\rm{decay}} =\frac{ (1-e^{\frac{-L_1}{|D_z|}})}{e^{\frac{L_s}{|D_z|}}}\,,
\label{eq:prob_decay}
\end{equation}
where $L_s$ denotes the starting position of the detector from the IP along the z-axis, and $L_1$ is the effective length of the detector that 
the LLP traverses before it decays inside the detector. We present a schematic diagram for this forward 
detector in Fig.~\ref{fig:forward_detector}. Depending on  the collimation of the LLP with the beam axis ($\theta$), $L_1$ can be determined 
as follows:
\begin{eqnarray}
L_1 = 0.0, ~~{\rm when}~~\theta > \tan^{-1}\left(\frac{L_s}{R}\right) \\ \nonumber
\end{eqnarray}  
\begin{eqnarray}
 L_1 = L_d,  ~~{\rm when}~~ \theta < \tan^{-1}\left(\frac{L_s+L_d}{R}\right) \,, \\  \nonumber
   \end{eqnarray}                
 \begin{equation}
 {\rm  else}~~L_1 = \frac{R}{\tan\theta} - L_s ,
\end{equation}      
where, $R$ is the radius of the detector, $L_d$ refers to the detector's length, $D_z=\frac{p_z}{m_\phi} c\tau$ is the distance traversed in the 
$z$-direction, where $\frac{p_z}{m_\phi}$ is the boost of the LLP in the $z$-direction, with $p_z$ being the momentum in the $z$-direction. 

Using Eq.~(\ref{eq:prob_decay}), we can calculate the acceptance of the signal event inside the detector using the formula:
\begin{equation}
\epsilon_{\rm LLP} = \frac{\sum_i P_{\rm decay}^{i}}{N_{\rm events}}\,,
\label{eq:llp_eff}
\end{equation}
where $i$ runs over all the long-lived particles in the simulated sample, and $N_{\rm events}$ represents the total number of events being analyzed. 
As discussed in the dark Higgs boson scenario in Section~\ref{s:model}, the long-lived scalar interacts with the SM-like Higgs boson, and the lifetime of such a scalar is inversely proportional to the square of the mixing angle $\mathrm{sin}\,\theta$. We follow Fig.~10 in~\cite{FASER:2018eoc} and map out the $c\tau$ corresponding to the respective values of sin\,$\theta$ and $m_{\phi}$. We further calculate the signal acceptance\,\footnote{Signal acceptance is the probability that the LLP decays inside the detector volume and has momentum greater than the required threshold of the detector. We have simulated $\phi$ production from $B^\pm$ meson decays for calculating the signal acceptance. The acceptance shall remain the same for $\phi$ coming from $B^0$ decays. We calculate the final sensitivity using the inclusive $B$ meson branching ratio as shown in Eq.\,\ref{eq:Ndetector}.} for the various combinations of $m_{\phi}$ and sin $\theta$ using Eq.~(\ref{eq:llp_eff}) to trace out the FASER2 sensitivity and validate our current analysis setup.
The number of events that decay inside the detector is given by:
\begin{equation}
N_{\rm detector} = \sigma_{bb} \times {\cal B}r(B \to \! X_s \phi) \times \epsilon_{\rm LLP} \times    {\cal L} 
\label{eq:Ndetector}
\end{equation}
where 
${\cal L} $ is the total integrated luminosity. To validate our result with the work by the FASER2 collaboration~\cite{FASER:2018eoc}, we 
have used the $B$-meson production cross-section $\sigma_{bb} = 9.4\times10^{11} \rm fb$~\footnote{Here we would like to 
mention that we have obtained the cross-section from \texttt{MadGraph} to be $\sigma_{bb}(NLO) = 3.99 \times 10^{11} \rm fb$. 
However, to validate the result with FASER collaboration, we have used the cross-section provided by Ref.~\cite{FASER:2018eoc}.}.
We assume that LLP decays to the visible particles with 100\% branching ratio and will be detected in the detector only if the 
momentum $p_{\phi} > 100\,$GeV~\cite{FASER:2018eoc}. 

\begin{figure}[htbp]
  \centering  
   \includegraphics[scale=1.3]{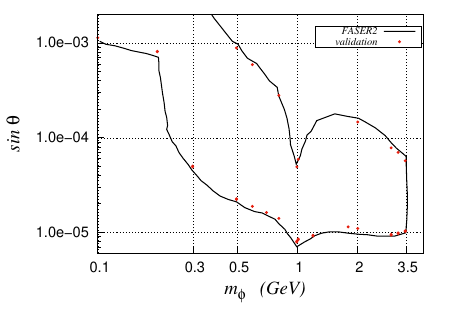}
   \caption{Validation of our analysis setup with FASER collaboration for dark Higgs scalar model. }
   \label{fig:valid_14}
  \end{figure}

In Fig.~\ref{fig:valid_14}, we show the sensitivity contour of the proposed FASER2 detector as a function of $\sin\theta$ and $m_{\phi}$, assuming 
that we observe three LLP decay events inside the detector volume, i.e., $N_{\rm detector}=3$,
and compare our results with the FASER2 collaboration. 
Inside the contour, we expect to observe more than three LLP events, and therefore, the detector will be sensitive in this region.
It is important to note that there could be an approximate uncertainty of 10\% while extracting the proper decay length from Fig.~10 in the reference paper~\cite{FASER:2018eoc}. Despite this uncertainty, we see from Fig.~\ref{fig:valid_14} that the projected sensitivity of FASER2 for various benchmark points of the dark Higgs model can be successfully reproduced, validating the current analysis setup. This ensures that the obtained results from our analysis setup are consistent with the expected sensitivity of the FASER2 experiment, providing confidence in the reliability of our analysis~\footnote{We have also validated our result for a few benchmark points with the FORESEE package~\cite{Kling:2021fwx}. Another framework for testing the sensitivity of multiple dedicated experiments is described in Ref.~\cite{Ovchynnikov:2023cry}.}.


\section{Dedicated Detectors at the 100 TeV FCC-hh}
\label{s:result}

In this section, we aim to motivate the need for new forward dedicated LLP detectors for the 100\,TeV FCC-hh collider experiment and explore their sensitivity. 
To begin with, let us examine the signal acceptance for  the long-lived dark Higgs boson {model at  
FASER2, as depicted in Fig.~\ref{fig:FASER14}. For the analysis, we consider benchmark points within a mass range of $m_\phi$ from 
0.1\,GeV to 4.4\,GeV and an independent 
variation of the proper decay length ($c\tau$) between from $10^{-2}$\,m to $10^4$\,m. 
This is useful in translating the results to any non-minimal dark Higgs model as well, where the mass and lifetime do not follow the particular relation as in the minimal dark Higgs model described in Section\,\ref{s:model}.
As we can 
see from Fig.~\ref{fig:FASER14}, the signal acceptances for the discussed benchmark scenarios are generally not high and attain a maximum 
value of $\sim 10^{-3}\%$ for moderate masses and proper decay lengths. Notably, for small proper decay lengths of the dark Higgs boson ($c\tau = 10^{-2}$\,m) 
and a mass greater than or equal to 1\,GeV, the signal acceptance drops significantly. This motivates us to investigate the sensitivity of the 
dark Higgs boson model using a forward detector instead of the FCC-hh. For simulating events at 100\,TeV, we generate $b{\bar b}$ events at 
the leading order (LO) using {\texttt pythia8}, and we use the \texttt{cteq6l1} PDF distribution for this purpose~\cite{Pumplin:2002vw}. However, for the computation of 
the projected sensitivity at 100\,TeV, we use the NNLO cross-section to be $\sigma_{bb}=3.09\times 10^{12}~\rm{fb}$ as given in Ref.~\cite{Mangano:2016jyj}.

\begin{figure}[hbt!]
  \centering
     \includegraphics[width=0.5\textwidth]{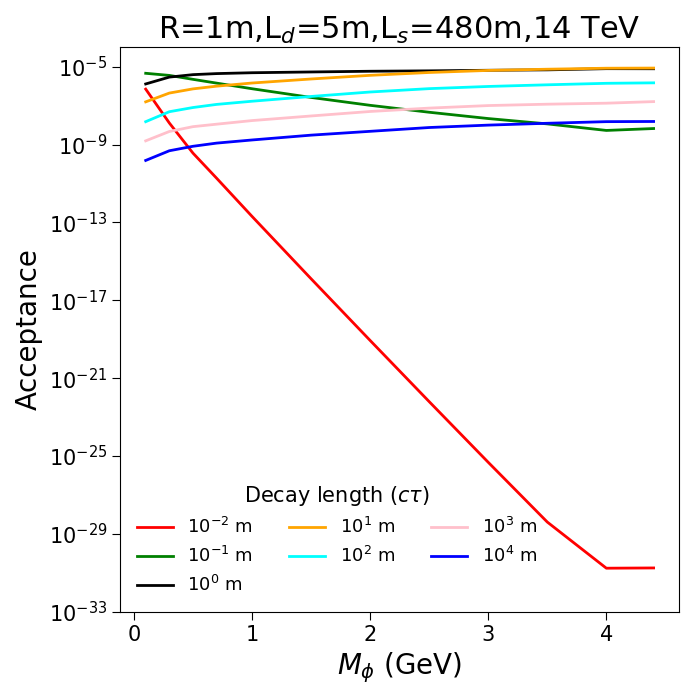}
     \caption{The signal acceptance for the dark Higgs scalar for proposed FASER2 detector at HL-LHC.
     }
  \label{fig:FASER14}
\end{figure}

\begin{figure}[hbt!]
  \centering
     \includegraphics[width=0.5\textwidth]{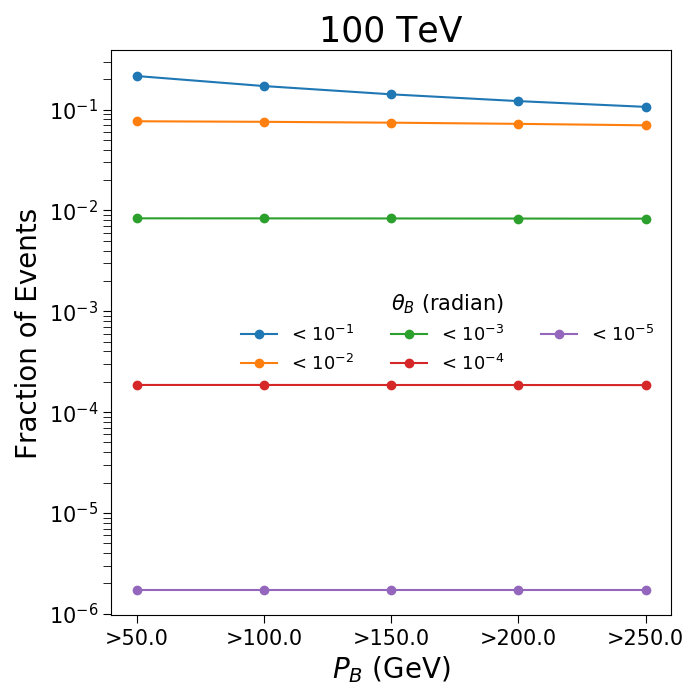}
   \caption{The fraction of surviving events after putting cuts on the production angle of the $B$-meson ($\theta_B$) and momentum of the $B$-meson ($p_{B}$) at the 100\,TeV FCC-hh collider.}
   \label{fig:ptheta}
  \end{figure}

At higher collision energies (100\,TeV), $B$-mesons can be produced in a more forward direction with higher momenta, which may 
enhance the chances of detecting long-lived scalar particles inside a forward detector, particularly for LLPs with small proper decay lengths 
and higher masses. However, we may need to apply a higher momentum cut to suppress the background that might be present at 
a 100\,TeV collider. Fig.~\ref{fig:ptheta} illustrates the fraction of events (expressed in percentages) that remain after implementing 
cuts on both the production angle ($\theta_B$) and momentum of the $B$-meson ($p_{B}$) at 100\,TeV. For FASER2, with $p_{B}> 
100$\,GeV and $\theta_B < 0.01$\,radian, the number of surviving events is 3.4\%. For this particular cut at 100\,TeV, the number 
of surviving events is 7.6\%,
implying an overall gain by a factor of 2. For a particular $\theta_B$ cut, the fraction of events remains constant even if we increase 
the $p_B$ cut. We also observe that if the $\theta_B$ cut is relaxed, i.e., the detector is placed near the IP or the radius of the 
detector is increased, the fraction of events can be increased further. 

In support of our above statement, we show the signal acceptance in Fig.~\ref{fig:faserlike100} as a function of the mass and $c\tau$ 
of the dark scalar with the same detector configuration and placement as FASER2 at 100\,TeV FCC-hh. Comparing signal acceptances 
for FASER2 at 14\,TeV and a FASER2-like configuration at 100\,TeV, we find considerable improvement for all benchmark points. Huge 
improvement for massive LLPs with very small proper decay lengths can be seen comparing Fig.~\ref{fig:FASER14} and 
Fig.~\ref{fig:faserlike100}. By examining the results, we observe that when $m_{\phi} = 0.1$\,GeV and $c\tau$ = $10^{-2}$\,m, the signal acceptance increases by a factor of 30. However, for $m_{\phi}=4.4$\,GeV and $c\tau$ = $10^{-2}$\,m, the signal acceptance increases 
by a huge factor of $10^{21}$. The reason behind such enhancement is due to the fact that the $B$-mesons produced in the 100\,TeV 
collider are very  energetic, and this high momentum is transferred to the $\phi$, giving a larger boost to the 
LLP. For this reason, $\phi$ can traverse longer distances and reach the detector despite having low $c\tau$. For example, we observe 
that with a cut of $p_{B} > 5 $\,TeV and $\theta_B < 0.01$\,radian, the number of surviving events at 14\,TeV is $10^{-4}\%$ whereas, for 
100\,TeV, this fraction increases to 0.4\%. 
  
  \begin{figure}[hbt!]
  \centering
  \includegraphics[width=0.5\textwidth]{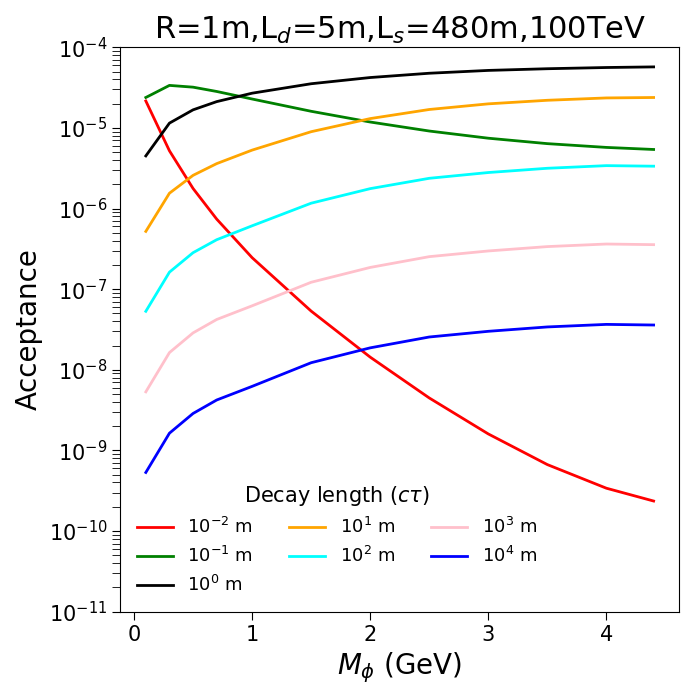}
  \caption{The signal acceptance for the dark Higgs model at 100\,TeV assuming a FASER2-like detector configuration. }
   \label{fig:faserlike100}
  \end{figure}

The placement of FASER2 inside the current LHC facility is constrained by the availability of space in the cavern. However, at future colliders like LHC at 100\,TeV, we are not limited by the space constraint since no particular decision regarding placement and general dimensions of a forward detector has yet been made~\cite{Mangano:2017tke,Mangano:2022ukr}. As an advantage, we have the freedom to carefully optimize the placement and size of the forward detector to get the maximum gain in signal acceptance. However, the placement and dimensions of the detector can be based on several other factors varying from the availability of space to the physics needs, signal gain versus cost ratio, and background yields.

\begin{figure}[hbt!]
   \begin{subfigure}{0.51\textwidth}
   \centering
 \includegraphics[width=\textwidth]{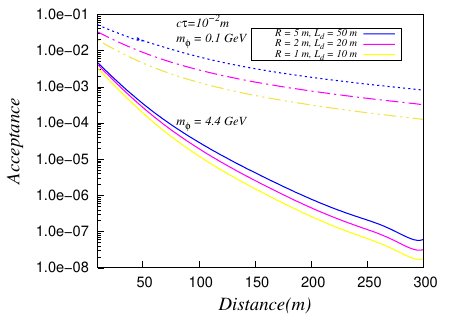}
 \caption{}
 \label{fig:4a}
   \end{subfigure}
   \begin{subfigure}{0.51\textwidth}
   \centering
 \includegraphics[width=\textwidth]{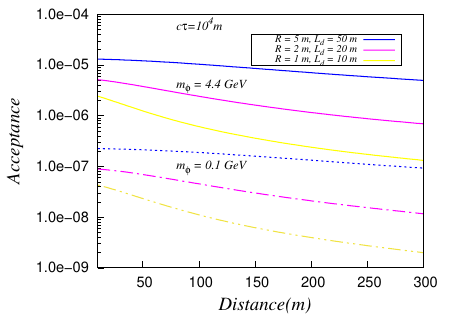}
 \caption{}
 \label{fig:4b}
 \end{subfigure}
   \caption{Variation of signal acceptance with the distance from IP for dark Higgs scalar model  for LLP masses 0.1 GeV(dashed line) and 4.4 GeV(solid line) and proper decay lengths of $10^{-2}$ m and $10^4$ m.}
   \label{fig:prob}
  \end{figure}

To explore the possible forward detector configurations for a 100\,TeV collider, we choose the four corner points of Fig.~\ref{fig:faserlike100} as benchmarks and study how signal acceptance varies as we change the detector position along the $z$-direction. We consider three different configurations for the detector, with varying values for the radius ($R$) and length ($L_d$). Specifically, we examine the 
configurations [$R$, $L_d$] $\in$ [1\,m, 10\,m], [2\,m, 20\,m] and [5\,m, 50\,m].
Fig.~\ref{fig:prob} shows the variation of signal acceptance for LLPs with $m_\phi=$ 0.1\,GeV (\textit{dashed line}) and 4.4\,GeV (\textit{continuous line}), and $c\tau= 10^{-2}$\,m (\textit{left}) and $10^{4}$\,m (\textit{right}) for various detector configurations as discussed above. We observe that the signal acceptance falls with increasing distance from the IP for each detector configuration and it attains its maximum value when the detector is positioned closest to the IP for all benchmark points. An increase in the detector's length and radius helps us achieve higher signal acceptance. For example, if we place the detector at $z = 50$\,m, for $m_{\phi}= 4.4$\,GeV and $c\tau = 10^{-2}$\,m, the gain in acceptance is of the order of 2 if we change $R$ from 1\,m to 5\,m and $L_d$ from 10\,m to 50\,m. For the same set of configurations, for the benchmark point with $m_{\phi}= 0.1$\,GeV and $c\tau = 10^{-2}$\,m, we can increase the acceptance by a factor of 3. For higher $c\tau$ values, the increase in signal acceptance is even more. For $c\tau = 10^4$\,m, the enhancement is by a factor of 10 for both masses of 0.1\,GeV and 4.4\,GeV. 
There is another interesting pattern to be noticed here. For small proper decay lengths, signal efficiency tends to decrease with increasing mass, while for larger proper decay lengths, signal efficiency tends to increase with increasing mass. 
The probability to decay inside the detector volume goes as Eq.~(\ref{eq:prob_decay}), where $D_z = \frac{p_z}{m_\phi} c\tau$. For larger values of $c\tau$, therefore, one needs smaller boost factors to increase this probability. This becomes true for heavier LLPs, which explains this observed trend.

The detector design for FCC-hh is not yet finalized.
For our work, we follow the detector placement guidelines outlined in Ref.~\cite{Mangano:2017tke,Mangano:2022ukr}, which suggests that the detector will be placed from $z\in[-25,25]$\,m. Keeping this in mind and assuming that a buffer length of 25\,m will be sufficient for shielding, we propose to place the forward detector at $z = 50$\,m. In case, it is not entirely feasible to veto the backgrounds with a 25\,m buffer length\,\footnote{The FCC-hh collider is expected to have higher background rates than the current LHC runs due to both an increased centre-of-mass energy of 100\,TeV and higher instantaneous luminosity. The specifications of the FCC-hh beam dimensions are not finalized or published yet to the best of our knowledge, which will affect how close to the IP we can place the detector. Therefore, a disclaimer is that the detector placement at 50\,m from the IP is subject to further studies of proper shielding materials to reduce the background at such a short distance and also the details of the beamline dimensions.}, we also provide two additional configurations with $z = 75$\,m and 100\,m, as listed in Table~\ref{tab:proposal}. We show a schematic visualization of the various proposed positions and geometries of the FOREHUNT and DELIGHT detectors at the FCC-hh in Fig.\,\ref{fig:forehunt_delight} for a better understanding of the readers.

\begin{figure}[hbt!]
  \centering
\includegraphics[width=0.95\textwidth]{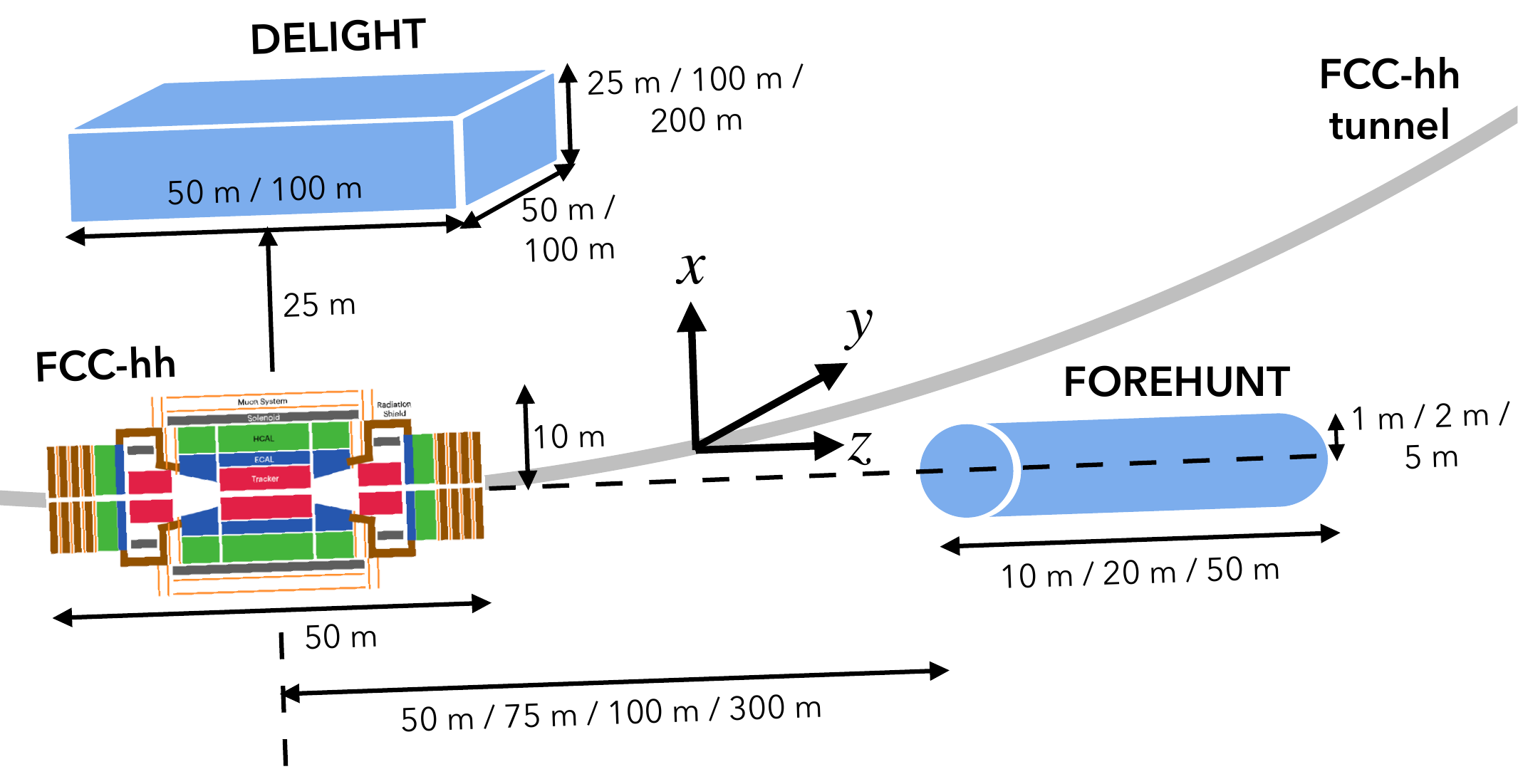}
\caption{Visualisation of the FOREHUNT and DELIGHT geometries at the FCC-hh.}
\label{fig:forehunt_delight}
\end{figure}

\begin{table}[hbt!]
\centering
\begin{tabular}{|l|l|l|l|}
\hline
\multicolumn{1}{|c|}{\begin{tabular}[c]{@{}c@{}}Detector \\ Configuration\\ @100 TeV\end{tabular}} &
  Radius (R) &
  Length $(L_d)$ &
  Position (Z)
  \\ \hline
FOREHUNT-A  & 1 m & 10 m & 50 m  \\ \hline
FOREHUNT-B  & 2 m & 20 m & 50 m \\ \hline
FOREHUNT-C  & 5 m & 50 m & 50 m  \\ \hline
FOREHUNT-D  & 2 m & 20 m & 75 m  \\ \hline
FOREHUNT-E  & 5 m & 50 m & 75 m  \\ \hline
FOREHUNT-F  & 5 m & 50 m & 100 m  \\ \hline
\hline
\end{tabular}
\caption{Proposed configurations for the dedicated LLP forward detectors at 100\,TeV, FOREHUNT.}
\label{tab:proposal}
\end{table}

Now, let us study how the signal acceptance changes with the change in the dimension of the detector, as depicted in
Fig.~\ref{fig:faser_lines_100} for six detector configurations as specified in Table~\ref{tab:proposal}, with $m_\phi$ varying between 0.1\,GeV to 4.4\,GeV and c$\tau$ varying between $10^{-2}$\,m to $10^{4}$\,m. In Fig.\,\ref{fig:faser_grid_100}, we show the signal acceptance grids for the six detector configurations in the $m_\phi$-$c\tau$ plane, for ease of translating our results to other models or reproducing our results.
From Fig.~\ref{fig:faser_lines_100}, we observe the following:

 \begin{figure}[hbt!]
  \centering
   \begin{subfigure}{0.33\textwidth}
   \centering
  \includegraphics[width=\textwidth]{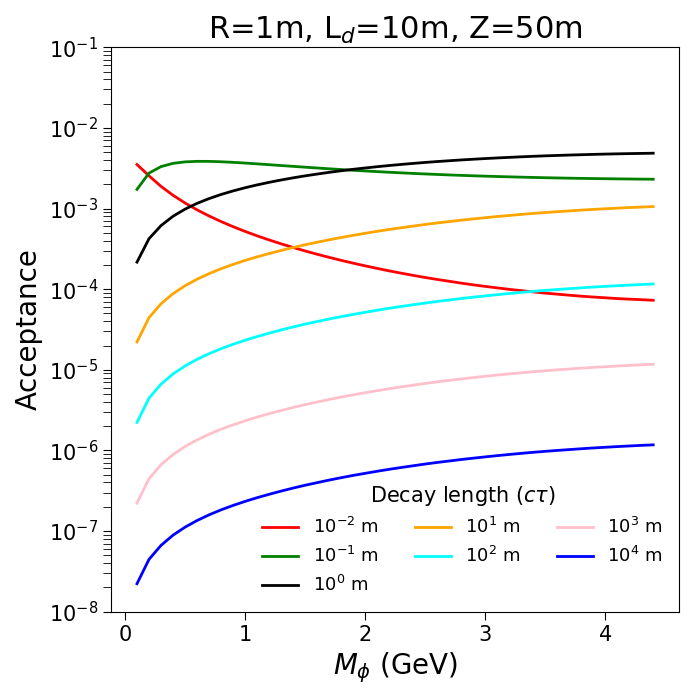}
  \caption{}
 \label{fig:3a}
  \end{subfigure}~
  \begin{subfigure}{0.33\textwidth}
  \centering
  \includegraphics[width=\textwidth]{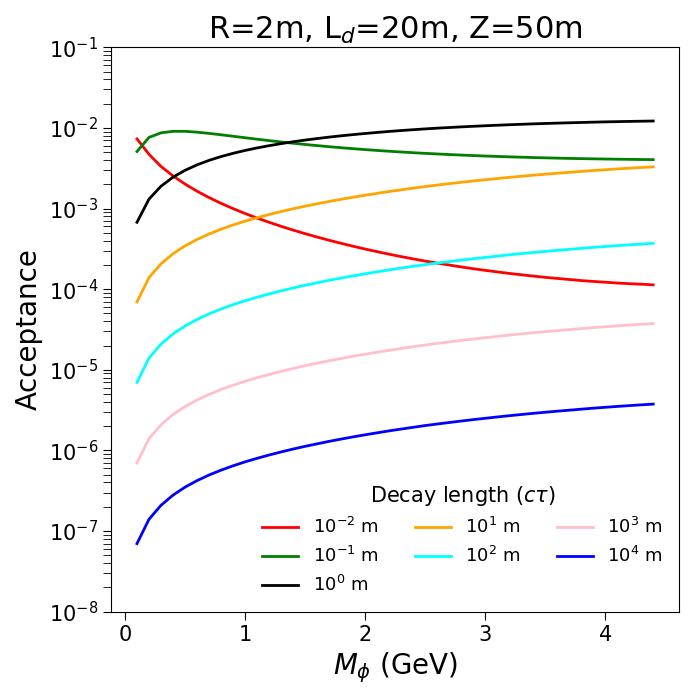}
  \caption{}
  \label{fig:3b}
  \end{subfigure}~
  \begin{subfigure}{0.33\textwidth}
  \centering
  \includegraphics[width=\textwidth]{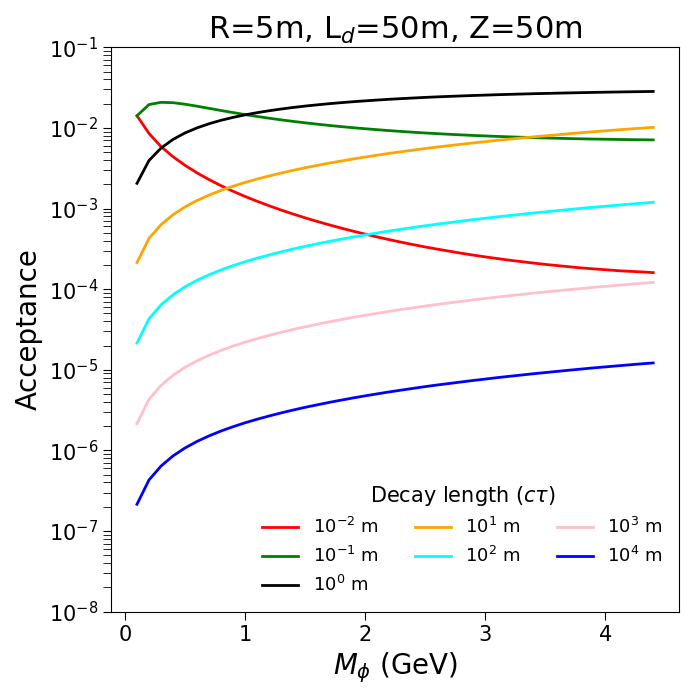}
  \caption{}
  \label{fig:3c}
  \end{subfigure}\\
  \begin{subfigure}{0.33\textwidth}
  \centering
  \includegraphics[width=\textwidth]{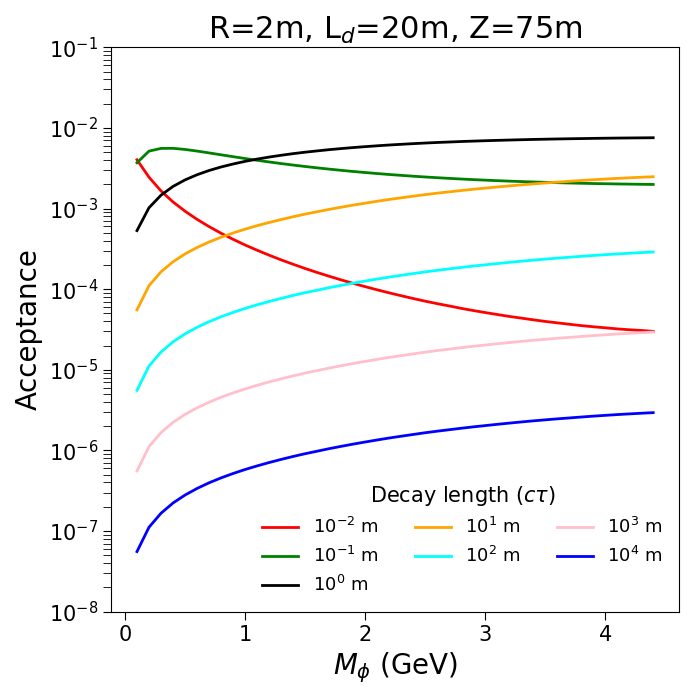}
  \caption{}
  \label{fig:3d}
  \end{subfigure}~
  \begin{subfigure}{0.33\textwidth}
  \centering
  \includegraphics[width=\textwidth]{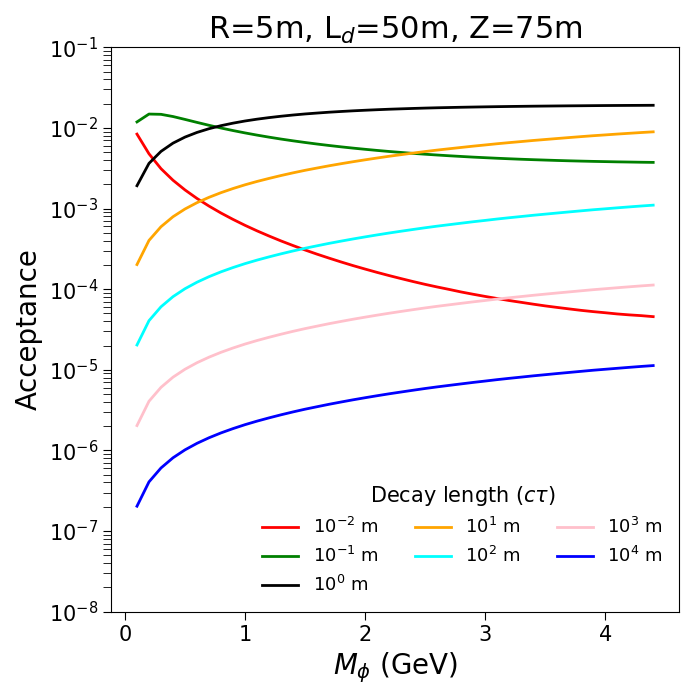}
  \caption{}
  \label{fig:3e}
  \end{subfigure}~
  \begin{subfigure}{0.33\textwidth}
  \centering
  \includegraphics[width=\textwidth]{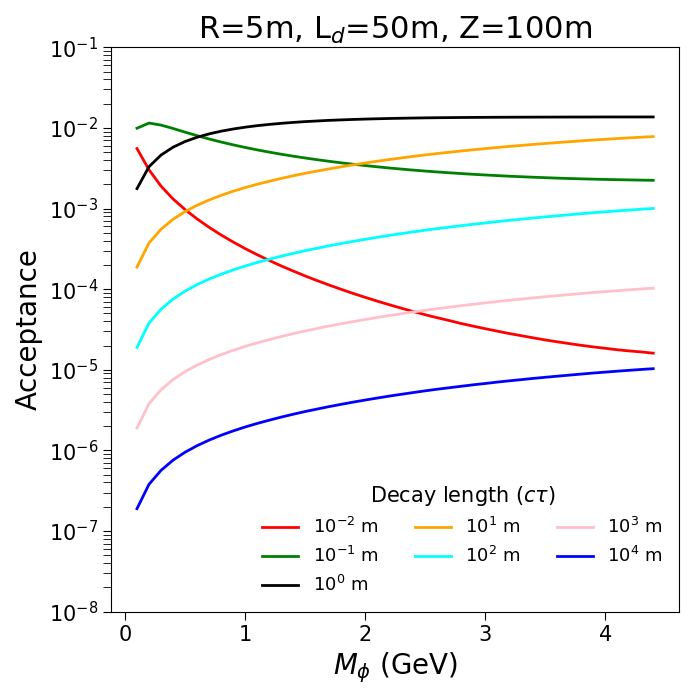}
  \caption{}
  \label{fig:3f}
  \end{subfigure}
   \caption{Signal acceptance for the dark Higgs model for different 100 TeV detector configurations as mentioned in Table~\ref{tab:proposal}.
   }
   \label{fig:faser_lines_100}
  \end{figure}

  \begin{figure}[hbt!]
  \centering
   \begin{subfigure}{0.5\textwidth}
  \includegraphics[scale=0.4]{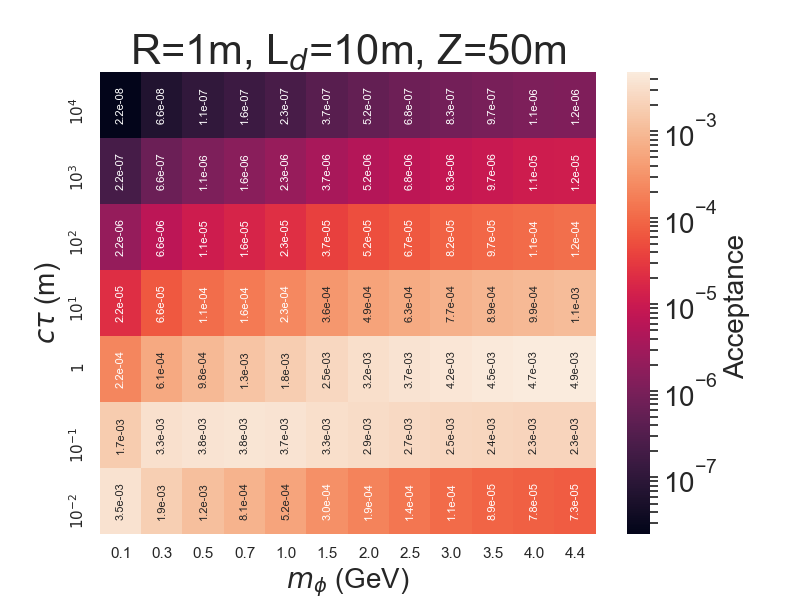}~~
  \caption{}
 \label{fig:3a_grid}
  \end{subfigure}
  \begin{subfigure}{0.4\textwidth}
  \includegraphics[scale=0.4]{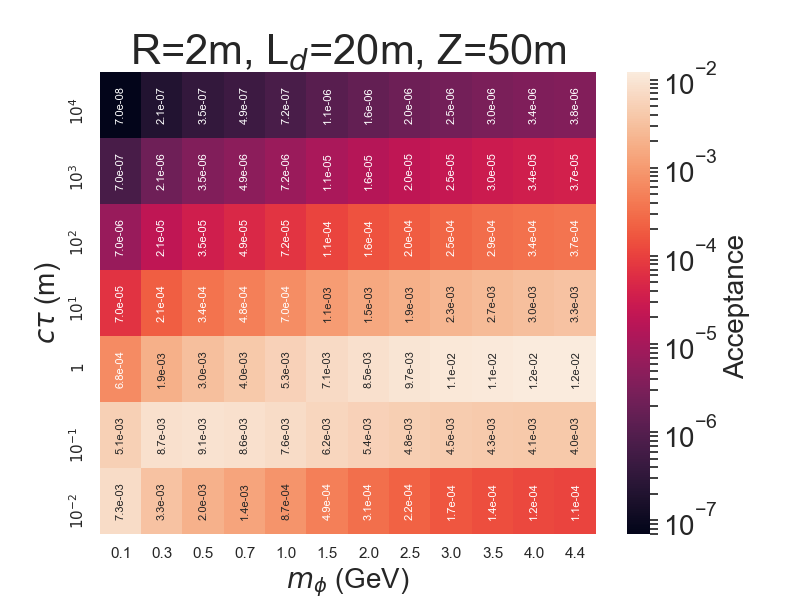}
  \caption{}
  \label{fig:3b_grid}
  \end{subfigure}
  \begin{subfigure}{0.5\textwidth}
  \includegraphics[scale=0.4]{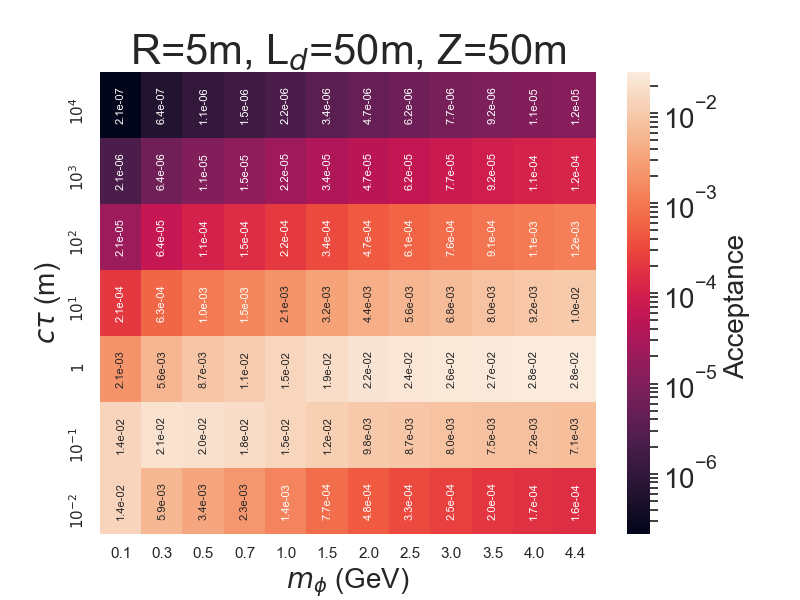}~~
  \caption{}
  \label{fig:3c_grid}
  \end{subfigure}
  \begin{subfigure}{0.4\textwidth}
  \includegraphics[scale=0.4]{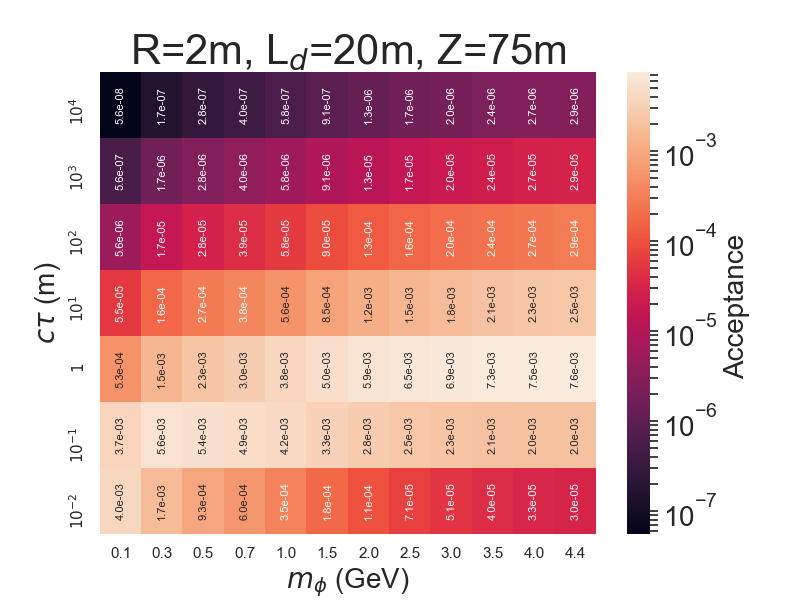}
  \caption{}
  \label{fig:3d_grid}
  \end{subfigure}
  \begin{subfigure}{0.5\textwidth}
  \includegraphics[scale=0.4]{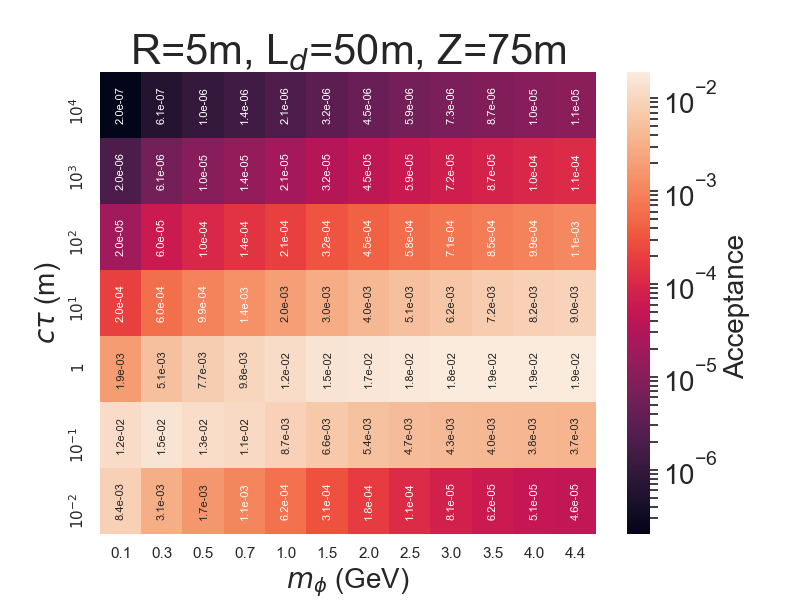}~~
  \caption{}
  \label{fig:3e_grid}
  \end{subfigure}
  \begin{subfigure}{0.4\textwidth}
  \includegraphics[scale=0.4]{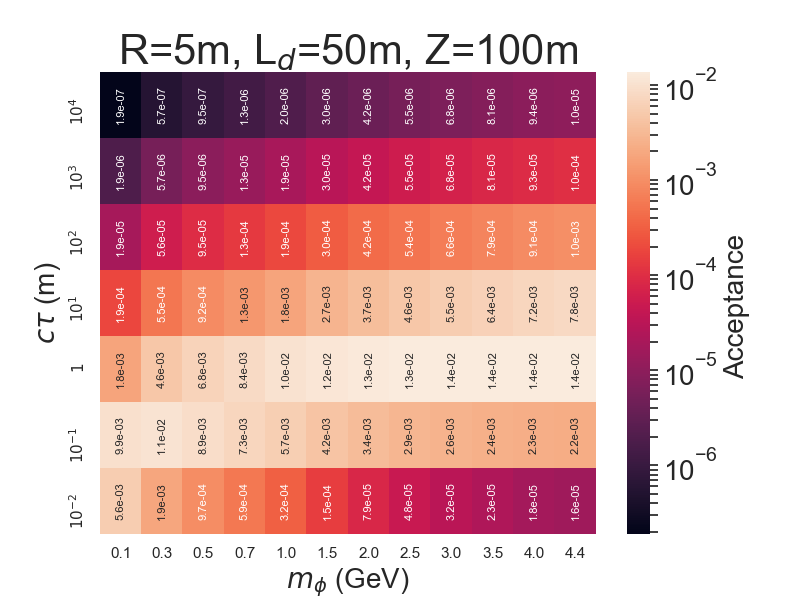}
  \caption{}
  \label{fig:3f_grid}
  \end{subfigure}
   \caption{Signal acceptance values for the dark Higgs model for different 100 TeV detector configurations as mentioned in Table~\ref{tab:proposal}.
   }
   \label{fig:faser_grid_100}
  \end{figure}
  
  \begin{itemize}
   \item We have seen earlier that compared to 14\,TeV (Fig.~\ref{fig:FASER14}), the signal acceptance at 100\,TeV increases drastically for high masses and low proper decay lengths (Fig.~\ref{fig:faserlike100}). We want to compare how the acceptance changes with our proposed detector configuration in Fig.~\ref{fig:3a}. We find that even for the light-dark Higgs with $m_{\phi}=0.1$\,GeV and $c\tau= 10^{-2}$\,m, the signal acceptance increases by a factor of 5000. 
   We gain in the signal acceptance by a factor of $\mathcal{O}(10^{26})$ for heavy LLPs ($m_\phi=4.4$\,GeV) with small proper decay lengths ($c\tau = 10^{-2}$\,m), where FASER2 at HL-LHC is not expected to be sensitive.
   For high $c\tau =10^4$\,m, the enhancement factor is 147 and 80 for masses 0.1\,GeV and 4.4\,GeV, respectively. 
   This massive improvement in signal significance is expected to be noticed at forward detectors at the FCC-hh.

   \item When the LLP proper decay length is small ($<1$\,m), the signal acceptance decreases as mass increases. We want to remind the 
   reader that this is due to the fact that the boost in the $z$-direction decreases with increasing mass, and since the proper decay 
   length is very small, heavy LLPs decay well before they reach the detector. This can be seen from Fig.~\ref{fig:3a}, as we go from 
   mass 0.1\,GeV to 4.4 GeV, for $c\tau=10^{-2}$\,m, the signal acceptance experiences a substantial reduction by a factor of 48. 
   
   \item For LLPs with considerable proper decay length ($>1$\,m), signal acceptance increases with mass because the probability of an 
   LLP to decay inside the detector increases with mass as can be seen from the Eq.~(\ref{eq:prob_decay}). For $c\tau= 
   10^4$\,m, the enhancement in signal acceptance is a factor of 54 if we change the mass from 0.1\,GeV to 4.4\,GeV.
   
   \item We note that increasing the radius of the detector can improve the signal acceptance for heavy LLPs. This is due to the 
   increase in the emission angle $\theta$, which necessitates a larger radius to capture the LLP decay. Moreover, increasing the 
   detector's length is also beneficial, as evidenced by Eq.~(\ref{eq:prob_decay}), which suggests that a longer detector 
   can result in higher probabilities of LLPs decaying inside the detector volume. For instance, for $m_{\phi} = 4.4$\,GeV and $c\tau=10^{-2}$m, if we change the radius from 1\,m to 5\,m and the length of the detector from 10\,m to 50\,m, comparing Fig.~\ref{fig:3a} and Fig.~\ref{fig:3c}, we observe the signal acceptance increases by a factor of 2.2. For many benchmark points, we can gain a factor of $\sim\!10$, for example, mass 1.0\,GeV and $c\tau= 10^1$\,m.
   
   \item It is evident from Fig.~\ref{fig:prob} and Fig.~\ref{fig:faser_lines_100} that a detector with a larger radius and length positioned closer to the IP will provide the highest signal acceptance. For example, comparing Fig.~\ref{fig:3c} and Fig.~\ref{fig:3f}, we can see that for low mass and low $c\tau$, the signal acceptance falls by a factor of 2 as we move the detector from $z = 50$\,m to $z = 100$\,m. For high mass and low $c\tau$, this reduction factor is $\simeq 10$.
  \end{itemize}
We observe from Fig.~\ref{fig:faser_lines_100} that a forward cylindrical detector with a radius of $R = 5$\,m and a length of $L_d = 50$\,m, which is placed at $z = 50$\,m from the IP (FOREHUNT-C), gives the maximum signal acceptance out of the six detector configurations for all the benchmark points. 
 
 \begin{figure}[hbt!]
  \centering
  \begin{subfigure}{0.51\textwidth}
 \includegraphics[width=\textwidth]{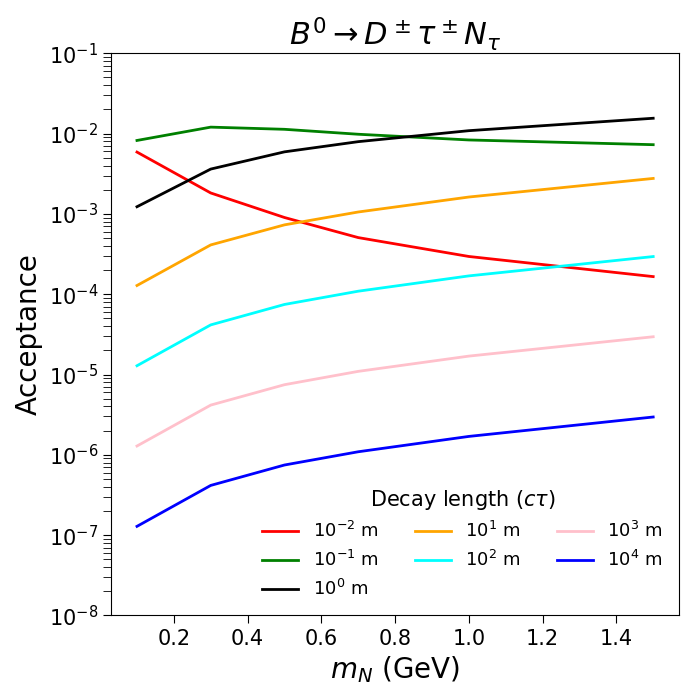}
 \caption{}
 \label{fig:7a}
 \end{subfigure}~
\begin{subfigure}{0.51\textwidth}
 \includegraphics[width=\textwidth]{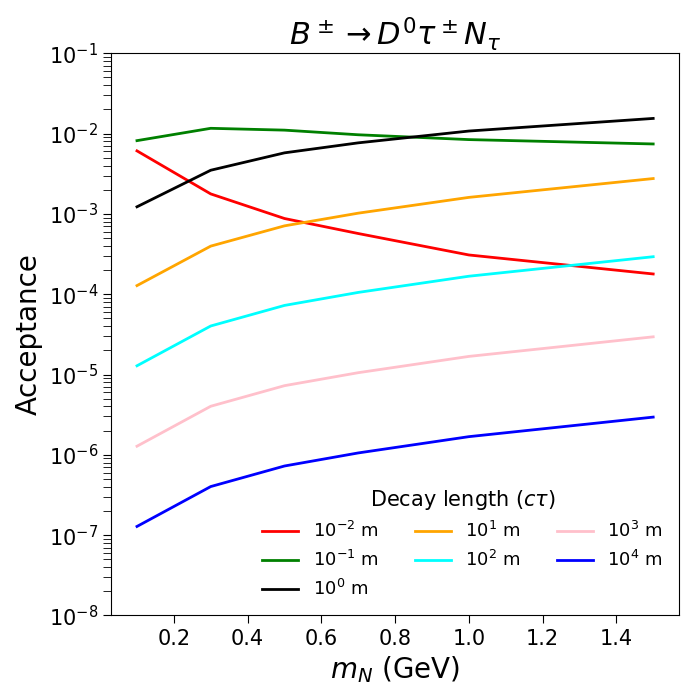}~~
 \caption{}
 \label{fig:7b}
 \end{subfigure}\\
 \begin{subfigure}{0.51\textwidth}
 \includegraphics[width=\textwidth]{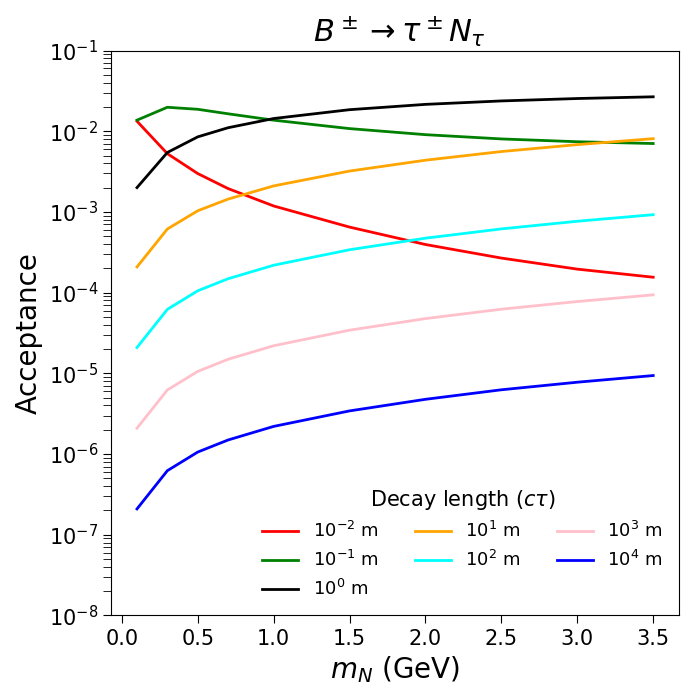}
 \caption{}
 \label{fig:7c}
 \end{subfigure}
 \caption{Signal acceptance for FOREHUNT-C detector configuration for heavy neutral lepton production. }
 \label{fig:HNL_effi}
 \end{figure} 
 
 Following a similar approach to the dark Higgs scalar scenario, we present the signal acceptance for HNLs [using Eq.~(\ref{eq:prob_decay})] with the FOREHUNT-C detector proposal in Fig.~\ref{fig:HNL_effi}. For this figure, we have assumed $100\%$ branching in each decay mode as discussed in Eq.~(\ref{eq:hnlproduce}).
 A similar pattern for signal acceptance is seen in Fig.~\ref{fig:HNL_effi} as for the dark Higgs boson model. For low $c\tau$, as we increase the mass of the HNL, the signal acceptance decreases, while for higher proper decay lengths, the signal acceptance improves as we increase the HNL mass. 
 
 Until now, we have only discussed the sensitivity of forward detectors. However, there are also a few proposed transverse detectors like MATHUSLA~\cite{Chou:2016lxi,Curtin:2018ees,Curtin:2018mvb,Alpigiani:2020iam} and CODEX-b~\cite{Gligorov:2017nwh,Aielli:2019ivi} at 14\,TeV, and DELIGHT at 100\,TeV~\cite{Bhattacherjee:2021rml}.
 Here, we would like to compare the results for FOREHUNT-C with these transverse detectors for six benchmark points with masses 0.1\,GeV, 2\,GeV and 4.4\,GeV, each having proper decay lengths of $10^{1}$\,m and $10^{4}$\,m. 
 In order to detect the LLP decaying inside such transverse detectors, it should have a minimum energy of $E_{\phi} > 1$\,GeV~\cite{Curtin:2018mvb}.
We now discuss the proposed locations and dimensions of these transverse detectors, and compare their performances with the forward detectors. 
We summarize the signal acceptances of FASER2, CODEX-b and MATHUSLA at 14\,TeV along with that obtained using the DELIGHT detectors and our proposal for FOREHUNT-C at 100\,TeV in Table~\ref{table:compare} for the six benchmark points mentioned earlier. 
The maximum signal acceptance for each benchmark point across all detectors is indicated in {\it bold font}.
We perform this comparison for the dark Higgs boson model in Table~\ref{table:compare}. Sensitivity studies of HNLs using transverse detectors like CODEX-b, ANUBIS, and MATHUSLA have been performed in Ref.\,\cite{DeVries:2020jbs}.

\underline{\bf CODEX-b:}

$\texttt{CODEX-b(``COmpact Detector for EXotics at
LHCb'')}$ is proposed to search  for LLPs decaying with $c\tau >$1\,m~\cite{Aielli:2022awh} at the  
HL-LHC with an integrated luminosity of 300\,fb$^{-1}$. This detector is 
to be installed near the LHCb interaction point using the existing infrastructure, making the costs low. The considered dimensions for this detector is a $10\times10\times10$\,m$^3$ decay volume and if possible, a bigger size of $20\times10\times10$\,m$^3$, with the following position:
\begin{itemize}
\item ${\bf{CODEX-b}}: 26.0\rm~ m < x < 46.0\rm~ m, -7.0\rm~ m < y < 3.0\rm~ m, 5.0\rm~ m < z < 15.0\rm~ m $
\end{itemize}
 
We find that CODEX-b outperforms FASER2 (Table~\ref{table:compare}) for all benchmark points considered here. 
However, with our proposed FOREHUNT-C, we see that even with high $c\tau >$1\,m, FOREHUNT-C will be more efficient. This increase is particularly notable for LLPs with shorter proper decay lengths. For instance, for an LLP with a mass of 4.4\,GeV and a proper decay length of 10\,m, the signal acceptance improves by $\sim 110$ compared to CODEX-b. Similarly, for a proper decay length of $10^4$\,m, the signal acceptance increases by a factor of $\sim 20$.
 
\underline{\bf MATHUSLA:}

$\texttt{MATHUSLA(``MAssive Timing Hodoscope for Ultra-Stable neutraL pArticles'')}$~\cite{MATHUSLA:2019qpy} is proposed to detect particle with $c\tau > 100$\,m at the 14\,TeV HL-LHC with an integrated luminosity of 3$ab^{-1}$.  MATHUSLA is intended to be positioned near the CMS interaction point at the HL-LHC.
The size of the detector is as follows:
\begin{itemize}
\item ${\bf{MATHUSLA}}: 60.0\rm~ m < x < 85.0\rm~ m, -50.0\rm~ m < y < 50.0\rm~ m,  68.0\rm~ m < z < 168.0\rm~ m $
\end{itemize}

Owing to its larger decay volume, MATHUSLA outperforms CODEX-b and FASER2 for all benchmark points, as presented in Table~\ref{table:compare}. However, for LLPs with a mass of  2\,GeV and 4.4\,GeV having a proper decay length of 10\,m, FOREHUNT-C significantly surpasses MATHUSLA in terms of signal acceptance. In the case of very low-mass LLPs, such as $m_\phi=0.1$\,GeV, MATHUSLA performs better than FOREHUNT-C. Also, for LLPs with a much longer proper decay length of $c\tau = 10^4$\,m, MATHUSLA demonstrates a relatively better performance as compared to FOREHUNT-C across all mass points.

\underline{\bf DELIGHT:}

$\texttt{DELIGHT(``Detector for long-lived particles at high energy of 100 TeV'')}$ is a transverse detector that is proposed to 
detect LLPs at the 100 TeV LHC, as discussed in ~\cite{Bhattacherjee:2021rml}. 
We have provided a visualisation of the DELIGHT geometries at the FCC-hh in Fig.\,\ref{fig:forehunt_delight}.
We calculate the signal acceptance for three proposed configurations given in 
Ref~\cite{Bhattacherjee:2021rml}:

\begin{itemize} 
\item ${\bf{DELIGHT-A}}: 25.0\rm~ m < x < 50.0\rm~ m, 0.0\rm~ m < y < 100.0\rm~ m, -50.0\rm~ m < z < 50.0\rm~ m $

\item ${\bf{DELIGHT-B}}: 25.0\rm~ m < x < 125.0\rm~ m, 0.0\rm~ m < y < 100.0\rm~ m, -50.0\rm~ m < z < 50.0\rm~ m $

\item ${\bf{DELIGHT-C}}: 25.0\rm~ m < x < 225.0\rm~ m,  0.0 m\rm~ < y < 50.0\rm~ m, -25.0\rm~ m < z < 25.0\rm~ m$
\end{itemize}

DELIGHT-A is proposed to be the same volume as MATHUSLA, DELIGHT-B is four times the volume of MATHUSLA, and DELIGHT-C is twice the volume of MATHUSLA.
Due to the large volume, DELIGHT-B gives the maximum signal acceptance out of the three configurations of DELIGHT. We find from Table~\ref{table:compare}, DELIGHT-B has the maximum signal acceptance for all the benchmark points except for $m_{\phi} = 4.4$\,GeV and $c\tau =10^{1}$\,m, where FOREHUNT-C has higher signal acceptance (by a factor of 4). However, for the benchmark with a higher proper decay length, DELIGHT outperforms FOREHUNT-C by a factor of $\gtrsim 15$ for all the chosen mass points.

\begin{table}[htbp!]
	\centering
	\resizebox{0.99\textwidth}{!}{
	\begin{tabular}{|p{1.5cm}|p{1.0cm}|c|c|c||c|c|}
		\hline
		
		$m_{\phi}$  & c$\tau$ & FASER2   & CODEX-b   & MATHUSLA &  FOREHUNT-C &  DELIGHT-B  \\  
		(GeV) & (m) &  ($p_{\phi}>100$GeV) & ($E_{\phi}>1$GeV)& ($E_{\phi}>1$GeV) & ($p_{\phi}>100$GeV) & ($E_{\phi}>1$GeV)  \\ \hline

          0.1 & 10$^{1}$ & 1.6 $\times 10^{-7}$  & 1.0 $\times 10^{-4}$ & 1.3$\times 10^{-3}$& 2.1 $\times 10^{-4}$ &  ${\bf{6.5 \times 10^{-3}}}$ \\ \hline

          0.1 & 10$^{4}$ & 1.5 $\times 10^{-10}$  & 1.1$\times 10^{-7}$ & 2.1$\times 10^{-6}$ & 2.1 $\times 10^{-7}$ & ${\bf{9.2 \times 10^{-6}}} $  \\ \hline

          2.0 & 10$^{1}$ & 3.6 $\times 10^{-6}$   & 1.8 $\times 10^{-4}$ & 4.4 $\times 10^{-4}$ & 4.4 $\times 10^{-3}  $ & $ {\bf{5.3 \times 10^{-3}}}$ \\ \hline

          2.0 & 10$^{4}$ & 4.8 $\times 10^{-9}$  & 1.9 $\times 10^{-6}$ & 3.4 $\times 10^{-5}$ & 4.7 $\times 10^{-6}$  &  ${\bf{1.5 \times 10^{-4} }}$   \\ \hline

          4.4 & 10$^{1}$ & 8.6 $\times 10^{-6}$  & 9.2 $\times  10^{-5}$ &  1.3 $\times10^{-4}$ & ${\bf{1.0\times 10^{-2}}} $ &  2.5 $\times 10^{-3}$ \\ \hline

          4.4 & 10$^{4}$ & 1.5 $\times 10^{-8}$  & 2.3 $\times 10^{-6}$ &  5.0 $\times  10^{-5}$ & 1.2 $\times 10^{-5}$ & ${\bf{1.9 \times10^{-4}}} $   \\ \hline

	\end{tabular}}
		\caption{Comparison of signal acceptance for detectors at 14 TeV (FASER2, CODEX-b and MATHUSLA) and at 100 TeV (FOREHUNT-C and DELIGHT) for six benchmark points for dark Higgs scalar model.}
	\label{table:compare}
\end{table}

 Table~\ref{table:compare} shows that DELIGHT performs very well 
 for most of the benchmark points with $c\tau > 10^1$\,m. We are keen to see how it performs for shorter proper decay lengths. In Fig.~\ref{fig:delight}, we present the signal acceptance of the three above-mentioned configurations of the DELIGHT detector in the $m_{\phi}-c\tau$ plane.
 By comparing Fig.~\ref{fig:6b} with Fig.~\ref{fig:3c}, we find that for LLPs with mean proper decay lengths less than 1\,m, the signal acceptance for FOREHUNT-C significantly exceeds DELIGHT-B, particularly for higher masses. For example, for $m_{\phi} = 4.4$\,GeV and $c\tau = 1$\,m, FOREHUNT-C outperforms DELIGHT-B by a factor of ${\cal{O}}(3\times 10^3)$. DELIGHT-B demonstrates superior performance for LLPs with proper decay lengths $\geq 10^2$\,m across all benchmark points. For intermediate proper decay lengths of around 10\,m, DELIGHT-B performs better than FOREHUNT-C for LLPs with masses $<2.5$\,GeV. However, for higher masses, FOREHUNT-C surpasses DELIGHT-B in terms of signal acceptance. For completeness, we have also given the signal acceptance for the proposed DELIGHT detectors with smaller sizes in Fig.~\ref{fig:6a} and Fig.~\ref{fig:6c}. We see for DELIGHT-A, the signal acceptance is reduced by a factor of 2 compared to DELIGHT-B for $c\tau \geq 10^2$\,m. For DELIGHT-C, the signal acceptance decreases by a factor of $\sim 2$ compared to DELIGHT-B for all masses and proper decay lengths. 
 This is a remarkable observation since it will have a great impact on the civil engineering costs of making the detectors. Although DELIGHT-C 
has a volume two times than that of DELIGHT-A, the former has a lower cross-sectional area towards the IP ($50\times50$\,m$^2$) as compared to DELIGHT-A ($100\times100$\,m$^2$). 
This increases the geometric acceptance of DELIGHT-A in the $\Delta\eta\times\Delta\phi$ plane.
So, if DELIGHT-B cannot be made because of its large size and volume, DELIGHT-A will be the next suitable transverse detector to be built for FCC-hh, instead of having a smaller volume than DELIGHT-C.

\begin{figure}[hbt!]
 \centering
 \begin{subfigure}{0.51\textwidth}
 \centering
   \includegraphics[width=\textwidth]{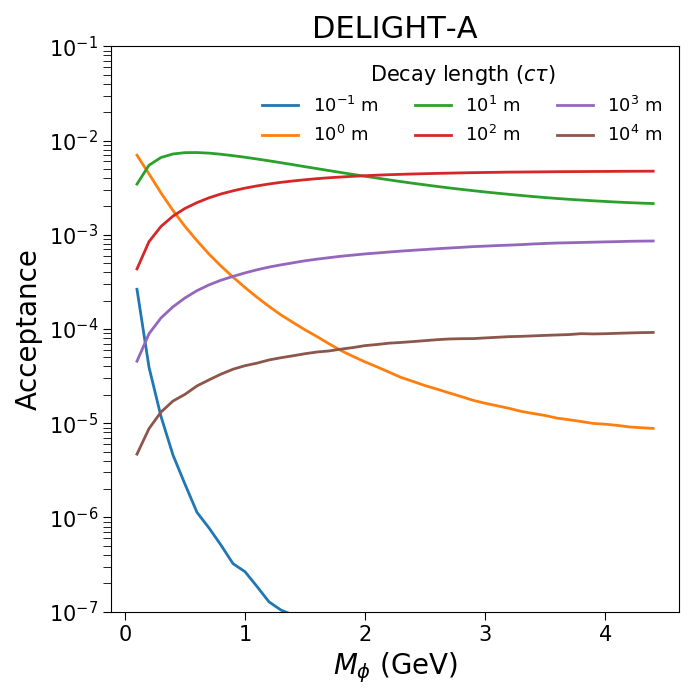}
   \caption{}
   \label{fig:6a}
   \end{subfigure}~
    \begin{subfigure}{0.51\textwidth}
    \centering
     \includegraphics[width=\textwidth]{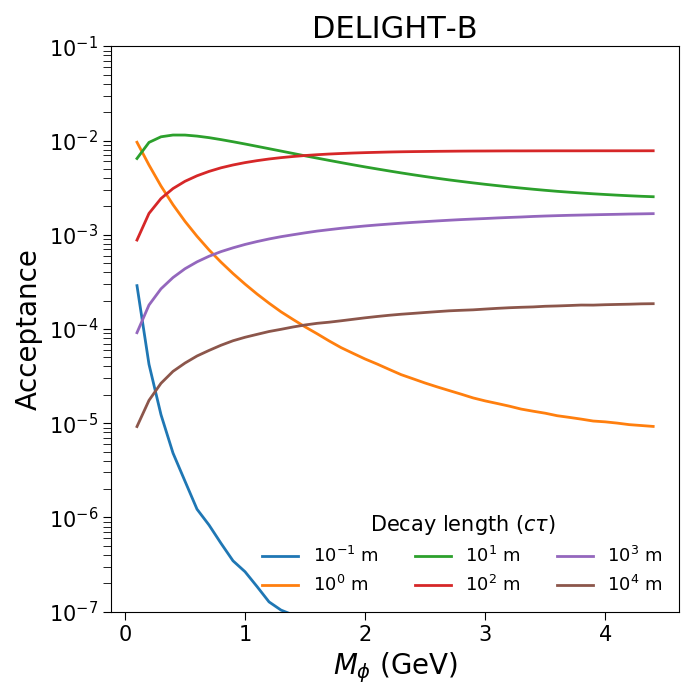}
     \caption{}
   \label{fig:6b}
     \end{subfigure} \\
      \begin{subfigure}{0.51\textwidth}
      \centering
   \includegraphics[width=\textwidth]{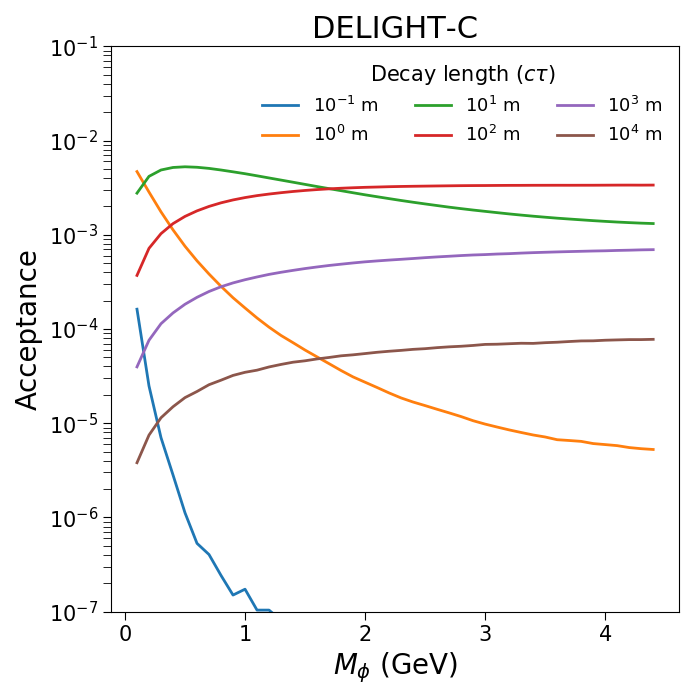}
   \caption{}
   \label{fig:6c}
   \end{subfigure}
   \caption{Signal acceptance for the transverse detector DELIGHT for the dark Higgs scalar model at the 100\,TeV FCC-hh.}
   \label{fig:delight}
  \end{figure}

Similar to the dark Higgs boson model, we want to investigate how DELIGHT performs for the heavy neutral leptons. 
Due to the larger decay volume, DELIGHT-B gives the maximum signal acceptance. We, therefore, depict the signal acceptance for this configuration as a function of mass and proper decay length of the HNL in Fig.~\ref{fig:delight_HNL}.
Comparing Fig.~\ref{fig:delight_HNL} and Fig.~\ref{fig:HNL_effi}, we find that when $c\tau > 10$\,m, DELIGHT-B performs better than FOREHUNT-C, similar to the dark Higgs model scenario~\footnote{From now on, we refer to the DELIGHT-B configuration as the DELIGHT detector, unless stated otherwise.}. 

\begin{figure}[hbt!]
 \centering
 \begin{subfigure}{0.51\textwidth}
 \centering
   \includegraphics[width=\textwidth]{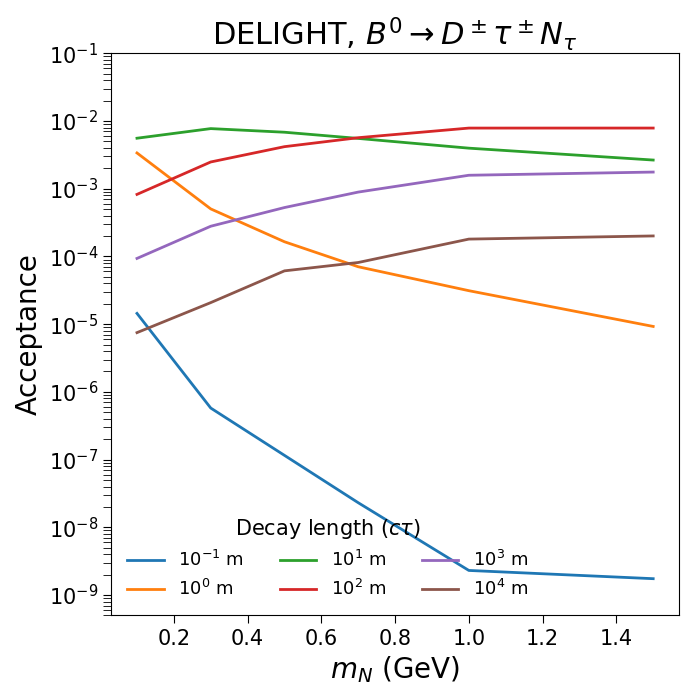}
   \caption{}
   \label{fig:6a_HNL}
   \end{subfigure}~
    \begin{subfigure}{0.51\textwidth}
    \centering
     \includegraphics[width=\textwidth]{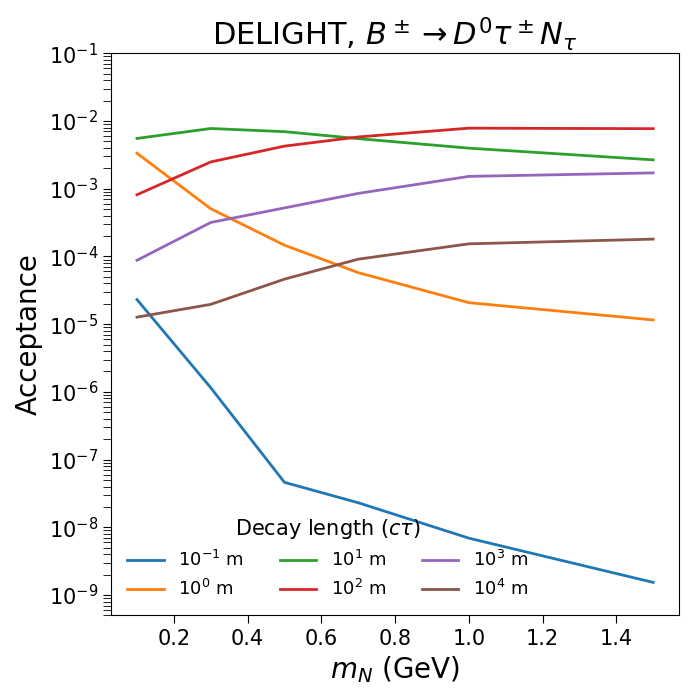} 
     \caption{}
   \label{fig:6b_HNL}
     \end{subfigure}\\
      \begin{subfigure}{0.51\textwidth}
      \centering
   \includegraphics[width=\textwidth]{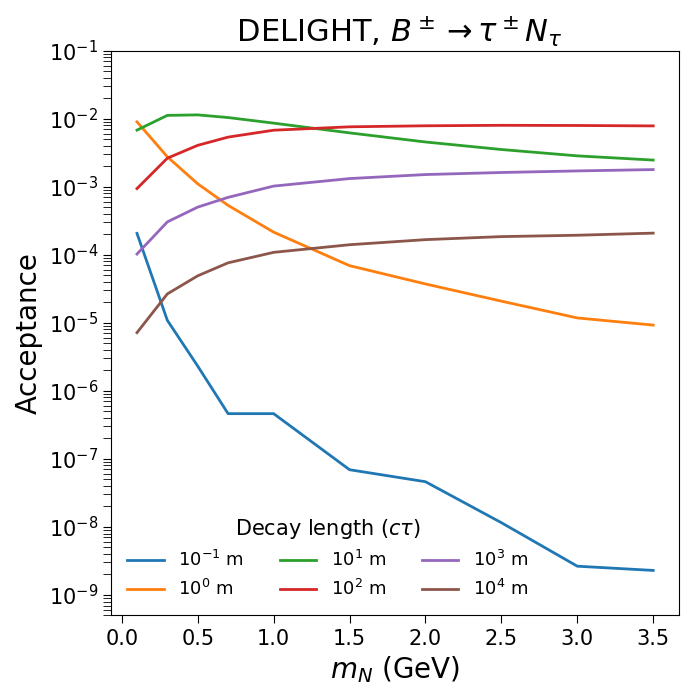}
   \caption{}
   \label{fig:6c_HNL}
   \end{subfigure}
   \caption{Signal acceptance for the transverse detector DELIGHT-B for heavy neutral leptons at the 100\,TeV FCC-hh.}
   \label{fig:delight_HNL}
  \end{figure}

Next, we investigate how such dedicated LLP detectors for the 100\,TeV FCC-hh improve the sensitivity in the model 
parameter space for the dark Higgs model. In Fig.~\ref{fig:thetaphi}, we translate the signal acceptance to the sensitivity in the plane of mixing 
angle $\sin\theta$ versus $m_{\phi}$, using Eq.~(\ref{eq:Ndetector}). Here we assume the integrated 
luminosity of the FCC-hh $\cal{L}$ $ = 30\,\rm ab^{-1}$ as an optimistic scenario. 
We see that compared to 14\,TeV, at 100\,TeV, 
we can improve the sensitivity in $\sin\theta$ by a factor of 15 and 20 for $m_\phi= 0.1$\,GeV and 1\,GeV, respectively, for the
FOREHUNT-C detector configuration. Moreover, we see that DELIGHT can enhance the sensitivity in $\sin\theta$ even more 
(factor of $\sim$ 5) compared to FOREHUNT-C. For the reader's interest, we also provide the relative reach for probing this 
model in various detectors like FASER2, CODEX-b, MATHUSLA, SHiP, FOREHUNT-C and DELIGHT in Fig.~\ref{fig:thetaphi}. 
We have extracted the sensitivity contours for CODEX-b, MATHUSLA, and SHiP from Ref.~\cite{FASER:2018eoc}. For $m_\phi<0.3$\,GeV, the parameter space is already constrained from existing experiments, like E949\,\cite{BNL-E949:2009dza} and NA62\,\cite{Lanfranchi:2671026}. For example, for $m_{\phi}=0.1$\,GeV, E949 has already probed a major part of the parameter space till a mixing angle of $10^{-4}$\,\cite{Ferber:2023iso}. 
However, with the FOREHUNT-C detector, we can improve this sensitivity and probe the mixing angle up to $\sim 6\times 10^{-5}$ and with DELIGHT-B, we can go further down to a mixing angle of $\sim 2\times 10^{-5}$.
Note that for DELIGHT-B, we only show the lower boundary to indicate the minimum mixing angles that can be probed by this detector.

\begin{figure}[hbt!]
\centering
\includegraphics[scale=1.5]{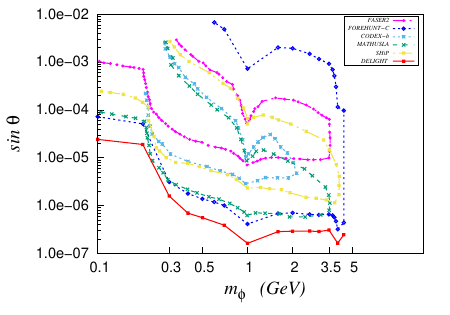}
\caption{The relative sensitivity reach for different detectors in the $\sin\theta$-$m_{\phi}$ plane for the dark Higgs scalar model.}
\label{fig:thetaphi}
\end{figure}
 
\begin{figure}[hbt!]
  \centering
\includegraphics[scale=1.5]{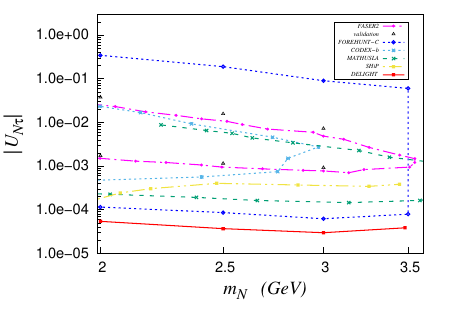}
\caption{The relative sensitivity reach for different detectors in the $|U_{N{\tau}}|$-$m_{N}$ plane for heavy neutral leptons.}
\label{fig:theta_compare_HNL}
\end{figure}
 
Similar to the dark Higgs model, we translate the signal acceptance into sensitivity to the mixing angle for the HNLs in Fig.~\ref{fig:theta_compare_HNL}. 
The sensitivity contours for CODEX-b, MATHUSLA, and SHiP for HNLs have been extracted from Ref.\,\cite{FASER:2018eoc}.
In the mass range 2\,GeV to 3.5\,GeV, HNLs can only be produced from the $B^{\pm}\rightarrow \tau^{\pm} N$ mode. We observe that, compared to FASER2, we can improve the sensitivity 
in the mixing angle by at least one order of magnitude in FOREHUNT-C. For DELIGHT, the enhancement is even more, by a factor of $\sim 6$ compared to FOREHUNT-C.

Before concluding, we would like to give attention to the following :
\begin{itemize}
 \item The NNLO $b\bar{b}$ production cross-section used for the 100\,TeV analysis can have an uncertainty of the order of 15\%~\cite{Mangano:2016jyj}. In addition, the choice of different PDFs can change the differential distributions of $B$-hadrons. To see this effect, we generate the parton-level $b$-pair events at the LO using different PDFs and have found that the signal acceptance varies at most $3\%$ for the dark Higgs model. However, these uncertainties do not noticeably change the sensitivity contours that we have obtained in Fig.~\ref{fig:thetaphi}.
 
 \item For future colliders, it is possible to place multiple forward detectors along the beamline. Observation of signals in multiple detectors would prove the presence of LLPs beyond any doubt. For the dark Higgs scalar model, we have investigated this possibility assuming four benchmark points mentioned in 
 Table~\ref{tab:multidetector}. Two detectors are placed at $z = 150$\,m or 300\,m in addition to FOREHUNT-C with the exact dimension of FOREHUNT-C, and we put a momentum cut $p_{\phi} >  100$\,GeV for detecting the LLP. We observe that the signal acceptance of the second detector decreases with an increase in the distance from the IP along the $z$-direction, as expected. Adding a second detector placed at $z = 150$\,m could potentially increase the signal acceptance by approximately 50\% compared to FOREHUNT-C. For higher proper decay length, even if we place the second detector at $z = 300$\,m, the overall signal acceptance increases by 50\%.

\begin{table}
\centering
\resizebox{0.99\textwidth}{!}{
\begin{tabular}{|p{1.5cm}|p{1.0cm}|c|c|c|}
\hline
$m_{\phi}$  & c$\tau$ & acceptance for & acceptance for & acceptance for \\
(GeV) & (m) & first detector at z=50 m &  second detector at z= 150 m &  second detector at z=300 m \\ \hline

0.1 & $10^{-1}$ & 1.4$\times 10^{-2}$ & 7.0$\times 10^{-3}$ & 2.9$\times 10^{-3}$ \\ \hline
4.4 & $10^{-1}$ & 7.0$\times 10^{-3}$ &  1.0$\times 10^{-3}$ & 1.9 $\times 10^{-4}$ \\ \hline
0.1 & $10^{4}$ & 2.1 $\times 10^{-7}$ & 1.6 $\times 10^{-7}$ & 9.3 $\times 10^{-8}$ \\ \hline
4.4 & $10^{4}$ & 1.2 $\times 10^{-5}$ & 8.5 $\times 10^{-6}$ & 4.9 $\times 10^{-6}$ \\ \hline
\end{tabular}
}
\caption{The gain in signal acceptance by putting multiple forward detectors instead of one detector for dark Higgs scalar model.}
\label{tab:multidetector} 
\end{table}

\item 
For the FOREHUNT detector, if it is not feasible to place the detector at 50\,m from the IP, we have to shift it farther away. For detectors placed far away from the IP, it might be possible to do better shielding and we wanted to study the effect of reducing the momentum threshold on the LLPs in such a case. 
We see from Table~\ref{table:alternate} that the signal acceptance with a selection cut of $p_{\phi} > 50$\,GeV for a detector starting from 100\,m is $\simeq 1.5$ times the signal acceptance when $p_{\phi} > 100$\,GeV cut (Table~\ref{table:compare}) is applied for a detector starting from 50\,m. 
In some cases, a forward detector with a lower energy cut can surpass the signal acceptance of DELIGHT(for example, $m_{\phi} = 2$\,GeV and $c\tau =10^4$\,m).

\begin{table}[htbp!]
	\resizebox{0.99\textwidth}{!}{
	\begin{tabular}{|p{1.5cm}|p{1.0cm}|c|c|c|}
	\hline
	$m_{\phi}$  & c$\tau$ & FOREHUNT-C & FOREHUNT-C & FOREHUNT-C \\
	(GeV) & (m) & ($p_{\phi} > 50$ GeV, z=100 m) & ($p_{\phi} > 50$ GeV, z=200 m) & ($p_{\phi} > 50$ GeV, z=300 m) \\ \hline
	 0.1 & 10$^{1}$ & 3.3$\times 10^{-4}$ & 1.8$\times 10^{-4}$  & 1.1$\times 10^{-4}$ \\ \hline
	 0.1 & 10$^{4}$ & 3.3$\times 10^{-7}$ & 1.8$\times 10^{-7}$  & 1.2$\times 10^{-7}$ \\ \hline
	 2.0 & 10$^{1}$ & 6.0$\times 10^{-3}$ & 3.0$\times 10^{-3}$ & 2.0$\times 10^{-3}$ \\ \hline
	 2.0 & 10$^{4}$ & 7.4$\times 10^{-6}$ & 4.4$\times 10^{-6}$ & 3.0$\times 10^{-6}$ \\ \hline
	 4.4 & 10$^{1}$ & 1.1$\times 10^{-2}$ &  5.0$\times 10^{-3}$ & 3.0$\times 10^{-3}$ \\ \hline
	 4.4 & 10$^{4}$ & 1.6$\times 10^{-5}$ & 9.0$\times 10^{-6}$ & 5.9$\times 10^{-6}$ \\ \hline
	\end{tabular}}
\caption{The signal acceptance for a detector like FOREHUNT-C with a smaller momentum cut for the dark Higgs scalar model.}
	\label{table:alternate}
\end{table}

Another possibility is to place multiple detectors in the forward direction as we discussed in the previous point, say one starting from 50\,m and a second detector at 200\,m. Since the first detector acts as an active veto for the second detector, the energy threshold for the latter might be reduced from 100\,GeV to 50\,GeV, assuming that the active veto helps to reduce the backgrounds.

\item Till now we have proposed forward detectors at 100\,TeV along the beamline similar to FASER2. However, the detector placement along the beamline might not be achievable or difficult for a detector like the FOREHUNT-C configuration being closer to the IP. We have studied the option of putting the forward detector slightly off the z-axis. We consider the FOREHUNT-C-like configuration, shifted in the transverse plane by 1\,m and 5\,m, and estimate the signal sensitivity for the dark Higgs model as listed in Table~\ref{tab:offzdetector}. As we place the detector slightly off-axis $(\sim 1$\,m), the acceptance drops by a factor of 2. If we place the detector at 5 m off-axis, the signal acceptance falls drastically. In such a case, instead of shifting the detector 5\,m in the transverse plane, it is more beneficial in terms of signal acceptance to place the detector farther away from the IP, say at $z = 300$\,m, along the beamline (see Table~\ref{tab:multidetector}).

\begin{table}
\centering
\begin{tabular}{|p{1.5cm}|p{1.0cm}|c|c|}
\hline
$m_{\phi}$ (GeV)  & c$\tau$ (m) & 1 m off-axis & 5 m off-axis \\ 
 & & ($p_{\phi} > 100$ GeV) & ($p_{\phi} > 100$ GeV) \\ \hline
0.1 & $10^{-1}$ & 8.3$\times 10^{-3}$ & 5.5$\times 10^{-4}$ \\ \hline
4.4 & $10^{-1}$ & 1.53 $\times 10^{-4}$ &  1.2 $\times 10^{-6}$ \\ \hline
0.1 & $10^{4}$ & 1.5 $\times 10^{-7}$ & 8.7 $\times 10^{-9}$ \\ \hline
4.4 & $10^{4}$ & 8.4 $\times 10^{-6}$ & 1.7 $\times 10^{-6}$ \\ \hline
\end{tabular}
\caption{The signal acceptance for putting the 100\,TeV FOREHUNT-C detector slightly off from $z$-axis for the dark Higgs scalar model.}
\label{tab:offzdetector} 
\end{table}

From Fig.~\ref{fig:thetaphi} and Fig.~\ref{fig:theta_compare_HNL}, one may come to the conclusion that a transverse detector like DELIGHT at 100\,TeV performs better than the forward detector, even for light LLPs. However, careful inspection of the signal acceptance grids given in Fig.~\ref{fig:3c_grid} and Fig.~\ref{fig:6b} shows that for $c\tau \lesssim 10^{1}$\,m, the forward detector performs comparably better than DELIGHT. 
This is clear from Fig.\,\ref{fig:forehunt_delight}, where we compare the signal acceptance of FOREHUNT-C and DELIGHT-B for $m_\phi=1$\,GeV. For lower decay lengths, the forward detector performs better than the transverse detector, and the trend remains the same across the LLP mass range considered here. 
Due to the larger boost in the forward direction at the hadron colliders, the forward detectors have better acceptance for lower decay lengths than the transverse ones for a particular LLP mass even when they are at similar distances away from the IP.
This is a model independent result and highlights the complementarity between the forward and transverse LLP detectors.
The upper boundary, corresponding to 3 observed events, for the DELIGHT detector requires simulating $\mathcal{O}(10^9-10^{13})$ events for the large mixing angles (small $c\tau$), which is beyond our computation limit. For the FOREHUNT detector, it's easier since we can use the probability formula without any assumptions due to the symmetry of the detector. We expect to observe at least 20 events at DELIGHT for a mixing angle of the order $2\times10^{-4}$, whereas for FOREHUNT, we expect to observe 20 events even for a mixing angle as large as $1.77\times10^{-3}$. This highlights that FOREHUNT is more sensitive for smaller $c\tau$ values.

\begin{figure}[hbt!]
\centering
\includegraphics[width=0.5\textwidth]{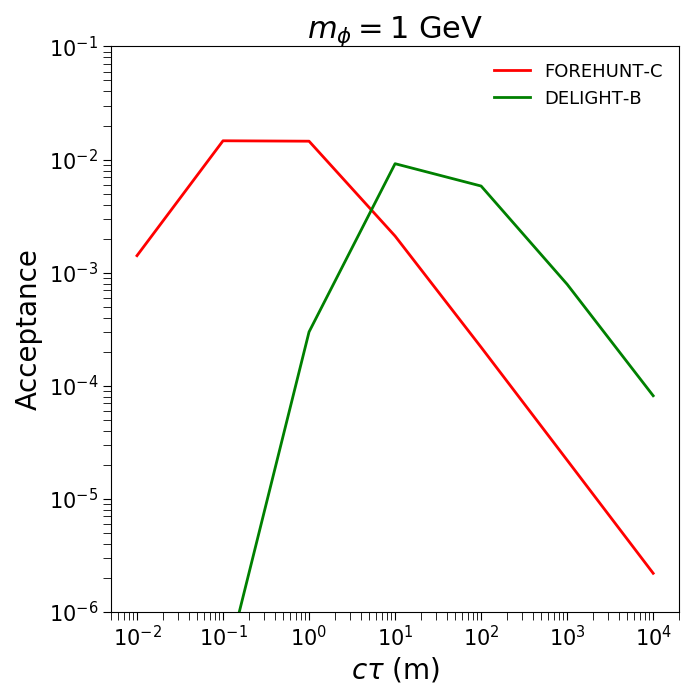}
\caption{Comparison between the signal acceptance of FOREHUNT-C and DELIGHT-B for an LLP mass of 1 GeV.}
\label{fig:forehunt_delight}
\end{figure}

\item For transverse detectors like 100\,TeV, we put the minimum energy cut of the LLP to be $>1$\,GeV as suggested by MATHUSLA~\cite{Curtin:2018mvb}.
For 100\,TeV, a 1\,GeV energy threshold might not be sufficient for triggering the LLP signal with acceptable background rates. Assuming a minimum energy threshold cut of 5\,GeV, we see that for low to moderate LLP masses (0.1\,GeV to 2\,GeV) the signal acceptance may decrease even by a factor of $\sim 7-8$, whereas, for heavier LLPs, a minor reduction of acceptance is observed.  
\end{itemize}

\section{Possible backgrounds and the detector design of FOREHUNT}

We have presented our sensitivity results assuming zero background events in the FOREHUNT detector. To understand how well this assumption works, we discuss the possible background sources in this section. Mitigation of these backgrounds also motivate the detector design and leads us to a possible design for the FOREHUNT detector.  
In the following list, we discuss the major background sources and how they can be reduced with the addition of various components in the detector design and event selection criteria for FOREHUNT at FCC-hh:
\begin{enumerate}
\item Muons from the IP: The muon flux from the FCC-hh IP is quite high where $\mathcal{O}(10^9)$ muons are expected during LHC Run 3 at FASER\,\cite{FASER:2022hcn}, which is expected to increase by at least two orders of magnitude at the FCC-hh. To mitigate the muon background, we would require to place scintillators before the decay volume in order to veto the muons. The inefficiency of each scintillator plane to detect a muon is below $10^{-5}$ for the veto system used in FASER\,\cite{FASER:2023tle}. Therefore, if we have at least four such scintillator planes, the inefficiency of the total veto system would be below $10^{-20}$, and the muon background at FOREHUNT would be negligible.

\item Neutral hadrons: Neutral hadrons coming from the IP can be a possible source of background. Being neutral, they cannot be vetoed by the scintillators, and long-lived neutral hadrons, like $K_L$, can then decay inside the decay volume, mimicking an LLP signal\,\cite{FASER:2023tle}. To reduce the neutral hadron background, we require some shielding material before the decay volume. Since the proposal is to place the FOREHUNT decay volume closer to the IP as compared to FASER, we require to select the shielding material in order to suppress most of the neutral hadrons background. Long-lived neutral hadrons, like $K_S$, can also originate from the interaction of the muons with the shielding material, which can then be vetoed using the signal of the parent muon in a scintillator placed before the shielding material.

\item Neutrinos: There is also a high flux of neutrinos from the IP in the forward direction\,\cite{Kling:2021gos}, which can interact with the material of the various detector elements to produce charged particles. If the neutrinos interact in the shielding material, the produced charged particles can be vetoed if we place scintillators between the shielding and the decay volume. In case the neutrino interacts after the veto system, the number of such events can be further suppressed by demanding at least two reconstructed tracks coming from the same vertex, where the reconstructed vertex lies within the decay volume. The neutrino background can be further reduced by a higher cut on the calorimeter energy deposit as shown in Ref.\,\cite{FASER:2023tle}, whereas in Fig.\,\ref{fig:ptheta}, we show that the signal is not affected by a higher energy threshold. Moreover, the detector elements can be designed with a minimum amount of material for suppressing the neutrino background.
\end{enumerate}

\begin{figure}[hbt!]
  \centering
\includegraphics[width=0.85\textwidth]{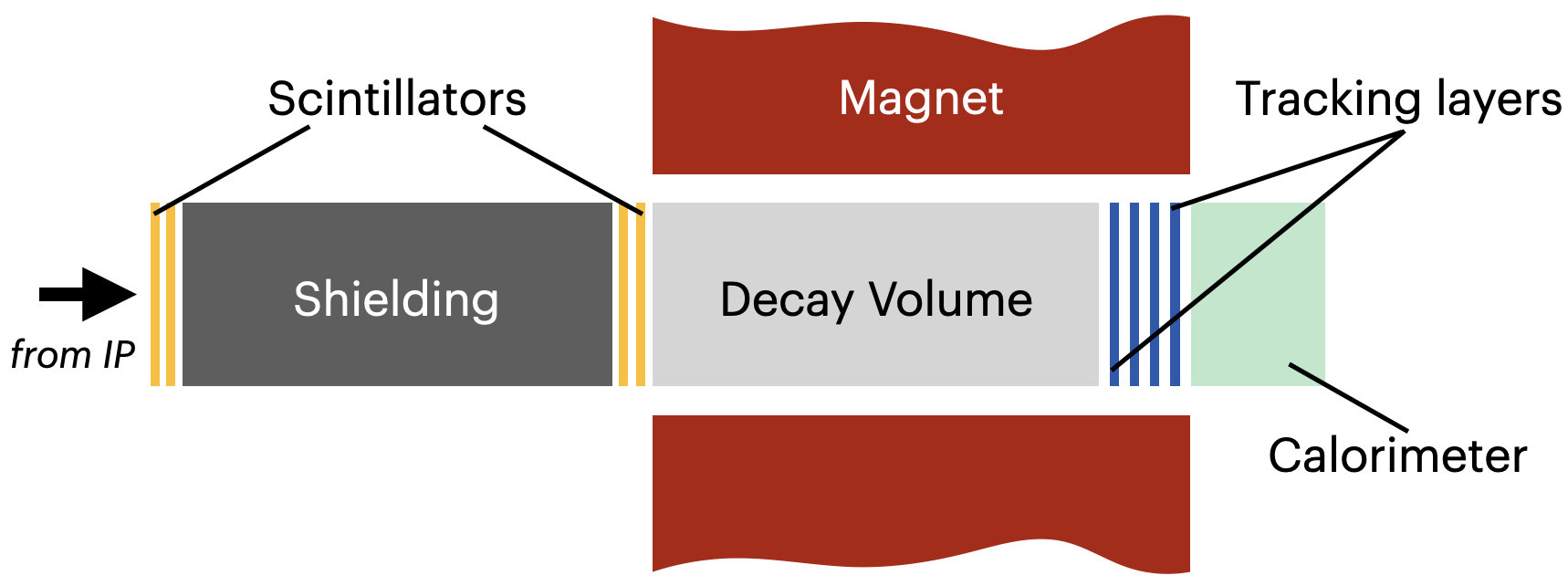}
\caption{Possible detector design for the FOREHUNT detector at FCC-hh.}
\label{fig:detector_design}
\end{figure}

\begin{figure}[hbt!]
  \centering
\includegraphics[width=0.48\textwidth]{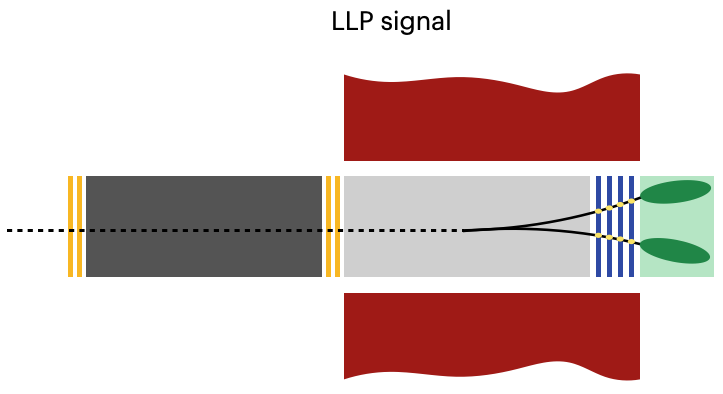}~~~
\includegraphics[width=0.48\textwidth]{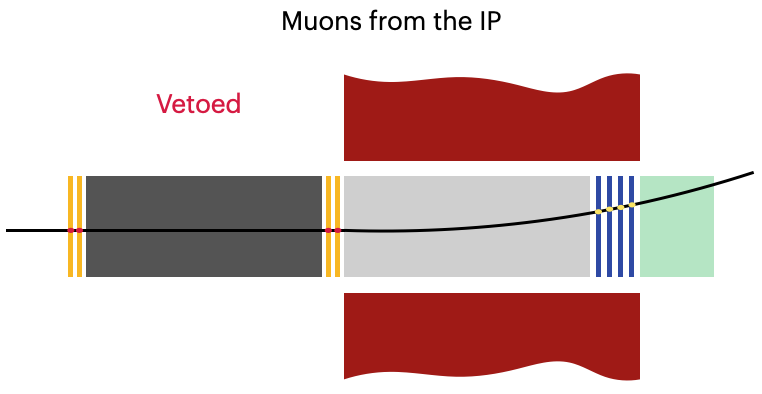}\\
\vspace{0.8cm}
\includegraphics[width=0.48\textwidth]{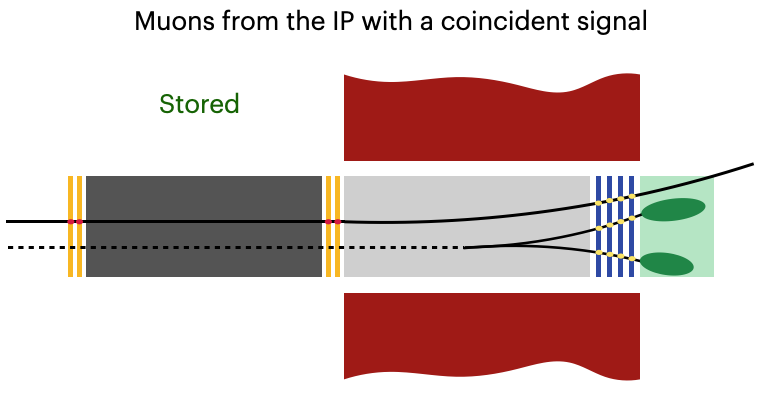}~~~
\includegraphics[width=0.48\textwidth]{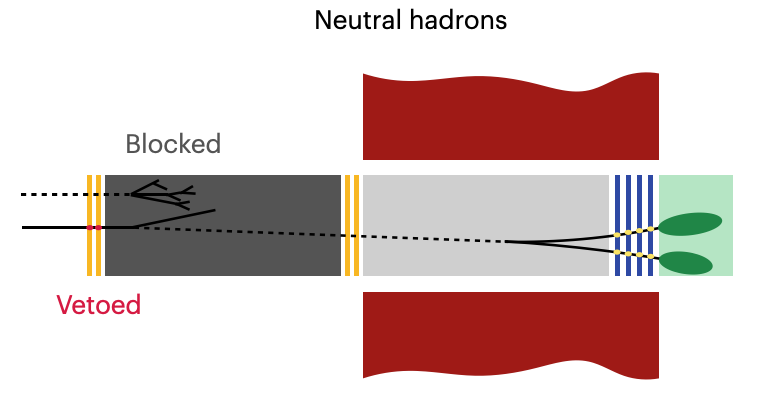}\\
\vspace{0.8cm}
\includegraphics[width=0.48\textwidth]{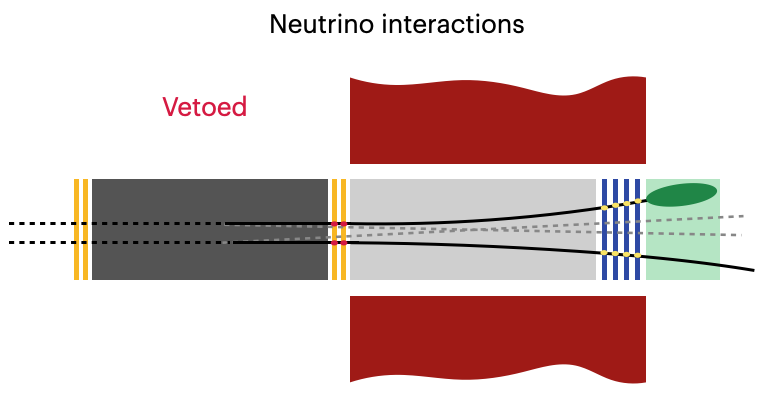}
\caption{Signatures of various processes at the proposed FOREHUNT detector $-$ an LLP decaying to electrons ({\it top left}), muons from the IP ({\it top right}), muons from the IP with a coincident signal ({\it center left}), neutral hadrons from IP or from muon interactions ({\it center right}), and neutrino interactions ({\it bottom}).}
\label{fig:sig_bkg}
\end{figure}

From the discussion above, we identify that the key components of the FOREHUNT detector should include $-$ at least four scintillator planes, a shielding block, a decay volume, tracking stations, and a calorimeter to detect the LLP decay products. Since the LLP will be highly boosted, its decay products will be collimated. A magnetic field is required to bend the decay products to produce isolated tracks and energy deposits. We require scintillator planes both before and after the shielding block to control the neutral hadron and neutrino backgrounds. Fig.\,\ref{fig:detector_design} shows a proposed design for the FOREHUNT detector and Fig.\,\ref{fig:sig_bkg} shows the signatures of the LLP signal and backgrounds from muons, neutral hadrons and neutrinos. This is a very preliminary outlook on the detector design and would require further dedicated studies in collaboration with experimentalists to optimize the detector materials, especially the shielding block. The proposal for the AL3X detector at IP2 of the LHC\,\cite{Gligorov:2018vkc} includes a 4\,m (40$\lambda$, where $\lambda$ is the nuclear interaction length) shield of tungsten and shows that it can suppress the primary hadron backgrounds to acceptable levels. At the FCC-hh collider, the hadronic background would be much higher compared to that at the LHC. Therefore, it would be important to identify materials that can block this huge background in the forward direction such that the LLP detector can be placed closer, increasing its sensitivity.

The veto system for the muons might affect the signal efficiency in two ways. The first is due to the noise in the scintillator planes which can give a false signal even when a MIP does not pass through it. This can lead to the veto of a signal event, reducing signal efficiency. However, the total noise of each sub detector for FASER falls within the range of $0.15 \pm 0.02$\,pC, which is negligible compared to a MIP signal peaking at $\sim 70$\,pC\,\cite{FASER:public_plots}. The FASER collaboration uses a threshold of 40\,pC on the scintillator charge for calculating the MIP efficiency\,\cite{FASER:public_plots}, which is much larger than the range of the noise distribution. Therefore, if we use similar quality of scintillator planes for the FOREHUNT detector, we expect manageable dark count rates.

The second possibility is the random overlap of a muon with the LLP signal. The FASER detector estimated this probability for their dark photon signal to be less than 1 per mille\,\cite{FASER:2023tle}. At FASER, the muon flux reaching the scintillators is around 650\,Hz. We expect this number to increase at the FCC-hh detector due to many factors: increased cross-section due to increasing centre-of-mass energy, increased instantaneous luminosity, and increased acceptance due to the closer placement of the detector. One idea to reduce this muon flux is to place a sweeping magnet in front of the detector to bend muons away from the detector. According to Ref.\,\cite{slides}, a sweeping magnet of 7\,Tm would bend 100\,GeV (1\,TeV) muons on the line of sight by 4.2\,m (40\,cm) from the line of sight, when the detector is 200\,m away from the sweeper magnet.

Another way to handle the coincidence of the muon with a signal event is to modify our veto system. We propose to do this by including the tracker and the calorimeter as parts of the veto system in addition to the scintillators. The modified veto condition is: veto events with hits in any of the four scintillator planes which do not have any calorimeter deposit other than ones which are consistent with the direction of the muon track inferred from the scintillator hits. 
The FCC-hh forward muon spectrometer can also play a role here in identifying the muon tracks which are within the FOREHUNT acceptance. We can then store events where the LLP signal is accompanied by muons. This might also store some background events where muons from the IP interact with the shielding material to produce a neutral hadron, which then decays within the FOREHUNT volume, given the neutral hadron is produced at a larger angle from the initial muon. However, a careful offline analysis of the event with the tracker hits or calorimeter energy deposits which are not consistent with the incoming muon can be done to differentiate such events from the signal. Sophisticated pattern recognition or other machine learning methods can be applied in the offline analysis. Exploring such possibilities in detail with proper detector simulations are left for a future study. Moreover, even if our signal acceptances are reduced by 50\%, Fig.\,\ref{fig:sigeff_50p} shows that there is only a slight change in the sensitivity reach of FOREHUNT-C for the two benchmark models studied here.

\begin{figure}[hbt!]
    \centering
    \includegraphics[width=0.5\textwidth]{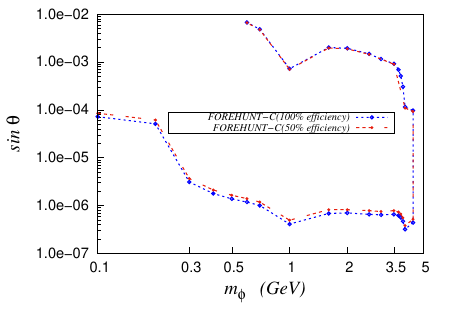}~
    \includegraphics[width=0.5\textwidth]{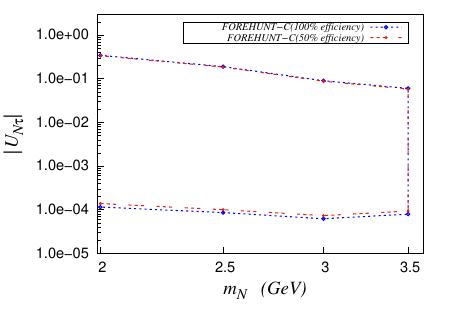}
    \caption{Effect on the sensitivity reach for the dark Higgs scalar ({\it left}) and the HNL ({\it right}) models with 50\% reduced signal acceptance.}
    \label{fig:sigeff_50p}
\end{figure}




For the tracking station, one can consider layers of Resistive Plate Chambers (RPCs).
The typical cost of BIS78 RPCs are around 3.1\,k\euro/m$^2$\,\cite{Bauer:2019vqk}. In our detectors, we propose to place several circular layers of RPCs transverse to the length of the cylindrical decay volume. Therefore, the cost of a single layer will be proportional to the cross-sectional area of the cylinder, which depends on the radius of the proposed detector.
For the maximum radius of 5\,m from our proposal, the cost per layer of RPC would be around 245\,k\euro. Increasing the number of such layers would increase the efficiency of reconstructing the secondary vertex. 
A triplet configuration of these RPC layers can achieve resolutions of 0.1\,cm and 0.4\,ns in the spatial and temporal directions, respectively\,\cite{Bauer:2019vqk}.
Moreover, the momentum resolution of the tracking system can be optimised by placing the tracking layers according to the Gluckstern equation\,\cite{tracking}, which relates the number of tracking layers, the distance between them and the magnetic field of the tracker to the momentum resolution of the reconstructed tracks.
The addition of a calorimeter will be useful for identifying LLPs decaying to photons, and it will help in extracting more information about the LLP, like its mass. Placing the detector close to the IP also opens up the opportunity to integrate it with the trigger system of the main FCC-hh detector.

\section{Conclusion}
\label{s:summary} 

In this paper, we study the detection prospect of light LLPs coming from $B$-meson decays at the FCC-hh. We consider two LLP candidates, 
namely a dark Higgs boson and a heavy neutral lepton. We propose a dedicated forward detector, named FOREHUNT, and study its sensitivity 
by optimizing the dimensions and placement of the detector. We see that in an optimistic scenario, if we could place the detector at 50\,m from the interaction point and with a radius of 5\,m, and length of 50\,m (FOREHUNT-C), the signal acceptance would be enhanced drastically as compared to the FASER2 detector for HL-LHC, mostly in high mass and low proper decay length region as discussed in Section~\ref{s:result}. There are two factors responsible for this enhancement. Firstly, compared to 14\,TeV, the $B$-mesons are much more energetic at 100\,TeV, and this boost is transferred to the LLP. Secondly, the forward detector is proposed to be placed near the IP, and the dimension of the detector is also large, resulting in an increase in the probability of decaying inside the detector. We estimate the future sensitivity of these models in terms of sensitivity to the mixing angle and mass of the LLP and see that compared to 14\,TeV, the reach of the forward detectors at 100\,TeV can enhance the sensitivity by a factor of 20 and $\approx 10$ for the dark Higgs model and HNL, respectively. We compare our result with another proposed detector at 100\,TeV, namely DELIGHT, and see that DELIGHT performs comparatively better than FOREHUNT-C for these two models. However, 
going beyond the minimal models,
we find that FOREHUNT-C performs better than DELIGHT when $c\tau < 10$\,m. We also discuss the comparison in sensitivity of the other transverse detectors, CODEX-b and MATHUSLA, at 14\,TeV.

The present work demonstrates the usefulness of forward-dedicated LLP detectors for the FCC-hh for light LLPs and their complementarity with the transverse detectors. We also discuss the prospects of placing multiple detectors along the beamline. We show that shifting the detector farther away from the IP is usually better than keeping it closer to the IP, but off-axis. To conclude, we encourage our experimental colleagues to consider these new proposals of the forward detectors at the FCC-hh that might contribute to the LLP search program.

  
\section*{Acknowledgement}
We thank Prof. Florian Bernlochner, Tobias B\"ockh and Sebastian Laudage for useful discussions regarding the FASER detector.
The work of BB is supported by the MATRICS Grant(MTR/2022/000264) of the Science and Engineering Research Board(SERB), Government of India. HKD and RS acknowledge the support of the Deutsche Forschungsgemeinschaft (DFG) through the funds provided to the Sino-German
Collaborative Research Center TRR110 “Symmetries and the Emergence of Structure in QCD” (DFG Project-ID 196253076)". NG would like to acknowledge the IOE-IISC fellowship program for financial support. SM is supported by Grant-in-Aid for Scientific Research from the MEXT, Japan (20H01895, 20H00153, 19H05810, 18H05542, JPJSCCA20200002) and by World Premier International Research Center Initiative (WPI), MEXT, Japan (Kavli IPMU). RS would like to thank the Indian Institute of Science for computational support.

\clearpage

\appendix

\section{LLPs from kaon decays}
\label{app:kaon}

The signal acceptance (\%) for dark Higgs scalars coming from the decay of kaon at the FCC-hh with the FOREHUNT-C detector is shown in Fig.\,\ref{fig:kaon}. 
We show the acceptances for two benchmark masses of 0.1\,GeV, and 0.3\,GeV. 
Comparing it with the corresponding signal acceptance obtained for dark Higgs scalars from $B$-meson decays, we observe that the larger proper decay length of kaons reduces the probability of the dark scalar to decay inside the proposed decay volume. 

\begin{figure}[hbt!]
    \centering
    \includegraphics[width=0.6\textwidth]{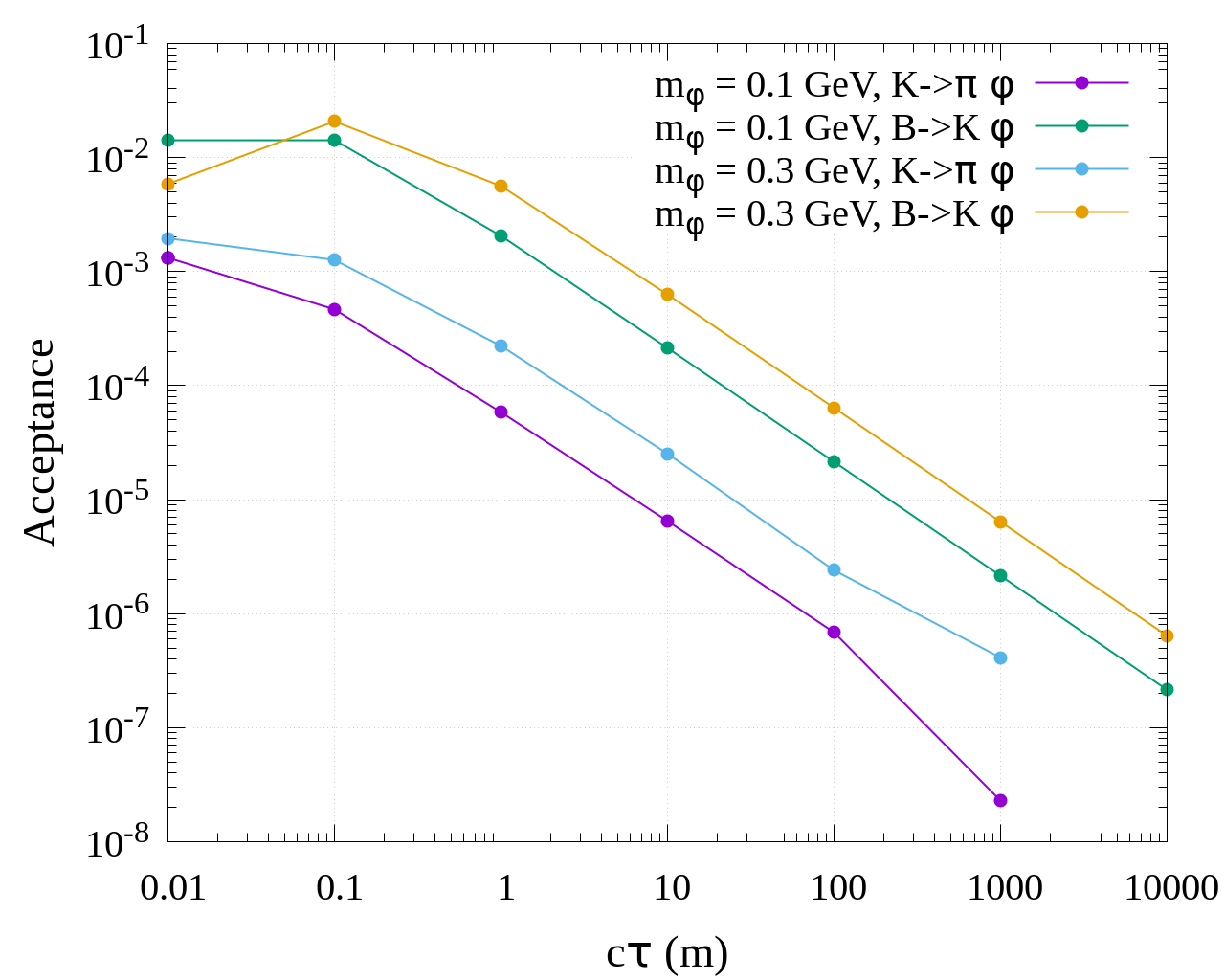}
    \caption{Variation of signal acceptance (\%) as a function of the proper decay length for two different dark Higgs scalar masses with the FOREHUNT-C configuration, when the dark Higgs scalar comes from the decay of a kaon. The corresponding efficiencies for dark Higgs scalars coming from $B$-meson decays are also shown for comparison.}
    \label{fig:kaon}
\end{figure}

\clearpage






%

\providecommand{\href}[2]{#2}\begingroup\raggedright\endgroup
\end{document}